\documentclass [11pt]{article}
\pdfoutput=1
\usepackage{amsfonts}
\usepackage{graphicx}
\usepackage{amsmath,amssymb, dsfont}
\usepackage{latexsym}
\usepackage{mathrsfs}
\usepackage{bbold}
\usepackage[dvipsnames]{xcolor}
\usepackage{slashed}
\usepackage{cancel}
\usepackage{soul} %
\usepackage{setspace} %
\usepackage{comment}
\usepackage{booktabs}
\usepackage{multirow}
\usepackage[nosort]{cite}
\usepackage[small]{caption}
\usepackage{array}
\usepackage[colorlinks,linkcolor=black,citecolor=blue,urlcolor=blue,linktocpage,pagebackref]{hyperref}
\usepackage{multirow}

\makeatletter
\def\section{\@startsection {section}{1}{\z@}{-3.5ex plus -1ex minus
 -.2ex}{2.3ex plus .2ex}{\large\bf}}
\def\subsection{\@startsection{subsection}{2}{\z@}{-3.25ex plus -1ex
minus -.2ex}{1.5ex plus .2ex}{\normalsize\bf}}
\makeatother
\makeatletter

\@addtoreset{equation}{section}

\makeatother

\def\nn{\nonumber}

\newlength{\diagh}
\allowdisplaybreaks

\newcommand\mt{{\widetilde m}}

\newcommand{\gt}{\widetilde g}
\newcommand{\xt}{\widetilde x}
\newcommand{\MSb}{$\mathrm{\overline{MS}}$ }

\DeclareMathOperator\arctanh{arctanh}
\DeclareMathOperator\arccot{arccot}

\begin{document}

\thispagestyle{empty}

\begin{center}

  \vspace*{-.6cm}

  \begin{center}

    \vspace*{1.1cm}

    {\centering \Large\textbf{Self-Dualities and Renormalization Dependence of the}}\\[3mm]

    {\centering \Large\textbf{Phase Diagram in 3d $O(N)$ Vector Models}}

  \end{center}

  \vspace{0.8cm}
  {\bf Giacomo Sberveglieri$^{a}$, Marco Serone$^{a}$, and Gabriele Spada$^{b,c}$}

  \vspace{1.cm}

  ${}^a\!\!$
  {\em SISSA and INFN, Via Bonomea 265, I-34136 Trieste, Italy}
  \vspace{.3cm}

  ${}^b\!\!$
  {\em Laboratoire Kastler Brossel, ENS - Universit\'e PSL, CNRS, Sorbonne Universit\'e,\\ Coll\`ege de France,  24 Rue Lhomond, 75005 Paris, France}
  \vspace{.3cm}

  ${}^c\!\!$
  {\em INO-CNR BEC Center and Dipartimento di Fisica, Universit\`a di Trento, 38123 Trento, Italy}

\end{center}

\vspace{1cm}

\centerline{\bf Abstract}
\vspace{2 mm}
\begin{quote}
  In the classically unbroken phase, 3d $O(N)$ symmetric $\phi^4$ vector models admit two equivalent descriptions connected by a strong-weak duality closely related to the one found by Chang and Magruder long ago. We determine the exact analytic renormalization dependence of the critical couplings in the weak and strong branches as a function of the renormalization scheme (parametrized by $\kappa$) and for any $N$. It is shown that for $\kappa=\kappa_*$ the two fixed points merge and then, for $\kappa<\kappa_*$, they move into the complex plane in complex conjugate pairs, making the phase transition
  no longer visible from the classically unbroken phase.
  Similar considerations apply in 2d for the $N=1$ $\phi^4$ theory, where the role of classically broken and unbroken phases is inverted.
  We verify all these considerations by computing the perturbative series of the 3d $O(N)$ models for the vacuum energy and for the mass gap up to order eight,
  and Borel resumming the series. In particular, we provide numerical evidence for the self-duality and verify that in renormalization schemes where the critical couplings are complex the theory is gapped. As a by-product of our analysis, we show how the non-perturbative mass gap at large $N$ in 2d can be seen as the analytic continuation of the perturbative one in the classically unbroken phase.
\end{quote}

\newpage

\tableofcontents

\section{Introduction}

Three-dimensional (3d)  $O(N)$ symmetric $\phi^4$ models are some of the most studied examples of non-integrable theories with interesting
RG flows. As well-known, by tuning the mass parameter these theories flow in the IR to a notable class of interacting conformal field theories.
Since in the IR these theories are strongly coupled, their exact RG flow is not known, though a lot can be said using a variety of techniques.

The appearance of a second-order transition is under perturbative analytic control within the $\epsilon$-expansion \cite{Wilson:1971dc}  for $\epsilon << 1$, and is natural to expect it to hold until $\epsilon=1$ (or $\epsilon=2$ for $N=1$), given the numerical evidence coming from Borel resummation techniques (assuming the series are Borel resummable).
The qualitative form of the perturbative RG flow of $O(N)$ models in the $\epsilon$-expansion in mass independent schemes is straightforward. For $m^2>0$ the theory is in the unbroken phase with a single gapped vacuum, the transition to a gapless phase occurs precisely at $m^2=0$, and for $m^2<0$ the theory is either gapped ($N=1$) or  gapless ($N>1$), with $O(N)$ symmetry spontaneously broken.
Crucially, we can directly study the critical theory at $m^2=0$ because of the absence of IR divergences within the $\epsilon$ expansion.
The gapless phase can also be accessed, for $2<d<4$, using large $N$ techniques.

Both large $N$ and $\epsilon$-expansion techniques are however not enough if one wants
to analyze the theory at finite $N$ within a non-perturbative definition of the theory, that requires to work with fixed integer dimensions.
At fixed dimension, IR divergences force us to work away from criticality with $m^2\neq 0$. In this case the gapless theory is defined when the physical mass gap $M^2$
(and not $m^2$)  vanishes. In a physical renormalization scheme, the gapless phase can in fact be reached
starting from $m^2>0$ by Borel resumming the perturbative series \cite{Parisi:1993sp}. This has been at the base of several
works for the extraction of critical exponents using resummation techniques, see e.g. \cite{Guida:1998bx}.
This is on a more firm footing with respect to the $\epsilon$-expansion, but it still cannot be considered a non-perturbative set-up, because
the physical renormalization scheme is only reached working order by order in perturbation theory \cite{Sberveglieri:2019ccj}.

In order to study the theory in a framework that can be compared with purely non-perturbative methods,  it is useful to work at fixed integer dimension and in ``minimal" renormalization schemes
where divergences are removed without the need of possibly inverting infinite perturbative series.\footnote{In fact, it is not a coincidence that the Borel resummability  proofs
  of \cite{Serone:2018gjo,Sberveglieri:2019ccj} (see also ref.\cite{eckmann1974,Magnen:1977ha}) apply only in such renormalization schemes.}
In $d=2$ several papers have indeed shown that the gapless phase can be reached in this way using lattice \cite{Milsted:2013rxa,Bosetti:2015lsa,Bronzin:2018tqz,Kadoh:2018tis},  Hamiltonian truncation \cite{Rychkov:2014eea,Elias-Miro:2017xxf,Elias-Miro:2017tup}\footnote{
  Hamiltonian truncations based on  light-cone quantization have also been used \cite{Burkardt:2016ffk,Anand:2017yij} but they require a non-trivial transformation to get mapped to the minimal
  covariant schemes we will discuss \cite{Fitzpatrick:2018ttk,Fitzpatrick:2018xlz}.} and Borel resummation \cite{Serone:2018gjo,Serone:2019szm} methods.

The aim of this paper is to extend to 3d Euclidean $O(N)$ $\phi^4$ models the study of the 2d $\phi^4$ theory of ref.~\cite{Serone:2018gjo}, where the unbroken phase was analyzed using Borel resummation
techniques of the perturbative series. We will not discuss in detail the properties of the critical theory, but rather focus more on how the phase diagram of the theory quantitatively depends on the choice of the minimal renormalization scheme and the appearance of self-dualities.

Renormalization scheme dependence is typically studied within perturbation theory, where couplings in different schemes are assumed to be related by an analytic mapping.
Within this approximation a phase diagram is invariant under coupling reparametrizations. For instance, given a scheme where a theory has fixed point $g^*$ with $\beta(g^*)=0$,
a change of scheme of the kind $g'=g'(g)$ gives
\begin{equation}
  \beta'(g') = \frac{dg'}{dg} \beta(g)\,.
\end{equation}
If the change of scheme is analytic, at $g^{\prime *}$ we would have $\beta'(g^{\prime *})=0$ and in particular the number of critical points would be in one to one correspondence in the two schemes.
But this is no longer true if we relax the assumption of analyticity of the coupling parametrization, since now $dg'/dg$ can have zeros or singularities on its own.\footnote{This is why we do generally care of higher order scheme-dependent terms of $\beta$-functions when studying the perturbative RG behavior of theories.
  Knowing the mere existence of a scheme where such terms vanish, and perturbative $\beta$-functions are saturated by one (or two) loop terms is useless, if we do not know
  the actual map $g'=g'(g)$.} This will be the case in our paper, where finite, non-perturbative, change of schemes are found and studied. However, we will not look for zeroes of resummed $\beta$-functions, but rather directly for points in parameter space where the mass gap $M^2$ vanishes.

We start in section \ref{sec:RSDSD} by setting the stage and defining the class of renormalization schemes we will consider, parametrized by the variable $\kappa$. By exploiting the super renormalizability of the theory, we show that 3d $O(N)$ models (in fact, we will consider at the same time both $d=2$ and $d=3$) admit two descriptions, equivalent to all orders in perturbation theory,
related by a strong-weak duality relation (within the same phase of the theory), closely related to a duality found by Chang and Magruder long ago \cite{Chang:1976ek,Magruder:1976px}.
Due to the Borel summability of the perturbative expansion, the relation is expected to hold at the non-perturbative level, at least in infinite volume on $\mathbb{R}^d$, the case we will consider.
Interestingly enough, the mass scale in the strong branch can be interpreted as a formal RG invariant dynamically generated mass scale of the theory in the weak branch.

The connection with the dualities found by Chang and Magruder, and the corresponding expected phase diagram for $N=1$, is described in section \ref{sec:Connec}, see fig.~\ref{fig:ChangRS}.
Using the results of section \ref{sec:RSDSD}, we analyze in section \ref{sec:FPAAD} the exact analytic dependence of the critical couplings in the weak and strong branches as a function of the renormalization scheme parameter $\kappa$, for any $N$. This is given by \eqref{eq:gcFunkappa} and is expressed in terms of the Lambert function $W$.
As $\kappa$ is varied in $d=3$, the two critical couplings change and move in opposite directions in the real coupling constant plane, until they merge at a given value $\kappa=\kappa_*$, after which they move into the complex plane in complex conjugate pairs, see fig.~\ref{fig:xplane}. When this happens, the phase transition
is no longer visible from the classically unbroken phase, if one restricts to {\it real} parameters in the Lagrangian.
On the other hand, the phase transition is always visible if one starts from the classically broken phase ($m^2<0$), where a single critical coupling is expected
for any value of $\kappa$. Based on these results, we speculate on a possible minimal analyticity structure of the $O(N)$ Schwinger functions in the complex mass plane, see fig. \ref{fig:Xana}.
In subsection \ref{subsec:LargeNMassGap}, as a by-product of our analysis, we show how the non-perturbative mass gap at large $N$ in 2d $\phi^4$ theories can be seen as the analytic continuation of the perturbative one in the classically unbroken phase.

In section~\ref{sec:Borel} we report our numerical results based on Borel resummations. As in our previous papers, we focus our attention on the 0- and 2-point functions.
In subsection \ref{subsec:coeff} we discuss how  we obtained  the perturbative coefficients up to order eight of the vacuum energy density $\Lambda$ and of the mass gap $M^2$ and how they are related in different renormalization schemes.\footnote{We define the mass gap $M^2$ as the zero of the Fourier transform of the inverse 2-point function (corresponding to the long-distance euclidean correlation length) and not as the pole of the propagator, as in our previous papers \cite{Serone:2018gjo,Serone:2019szm,Sberveglieri:2019ccj}.}
In subsection~\ref{sec:self-dual} we show the absence of a gapless phase for certain values of $\kappa<\kappa_*$, and provide evidence for the self-duality of the $O(N)$ models by
comparing the values obtained for $\Lambda$ and $M^2$ in the weak branch and (part of) the strong branch close to the self-dual point, see e.g. fig.~\ref{fig:Lambda_M2points}.\footnote{Evidence for a finite volume version of the self-duality of the 3d $N=1$ $\phi^4$ model has been recently provided using Hamiltonian truncation methods \cite{EliasMiro:2020uvk}.}
In subsection~\ref{sec:ReVEandM} we determine how the critical coupling in the weak branch moves as the renormalization scheme is varied, confirming the theoretical expectations,
and we compare the values of the critical coupling  with those obtained in the literature using lattice methods for $N=1,2,4$ \cite{Arnold:2001ir,Arnold:2001mu,Sun:2002cc}. The results are in fair agreement,
but with large uncertainties, due to the low accuracy of our resummations.
We conclude in section~\ref{sec:conclusions}. Four appendices complete the paper.
In appendix~\ref{sec:Lambert} we review some properties of the Lambert function $W$, which appears repeatedly in our analysis. In appendix~\ref{sec:LL4D} we compute $\Lambda$, $M^2$ and the zero momentum 4-point function for massive $O(N)$ models in $d$ dimensions at the first non-trivial order in the large $N$ limit. In appendix~\ref{app:MVEd3} we report the details of the vacuum energy renormalization in  3d $O(N)$ models. Finally in appendix~\ref{app:Coeff3d} we report the explicit form of the perturbative coefficients of $\Lambda$ and $M^2$ up to the eighth order in the coupling.

\section{Renormalization Scheme Dependence and Self-Dualities}
\label{sec:RSDSD}

In QFT the parameters entering the classical action, such as masses and coupling constants, do not correspond to physical observables
and are generally divergent. In the process of renormalization they get replaced by their renormalized and finite counterparts.
The precise definition of the renormalized parameters depends on the details of how we decide to renormalize the theory, i.e. from a renormalization scheme.
For definiteness, consider a dimensionless coupling $g$. If $g$ and $g'$ denote the renormalized coupling in two renormalization schemes,
in perturbation theory we have
\begin{equation}
  g' \sim g+ \sum_{n=2}^\infty a_n g^n\,,
  \label{eq:RS1}
\end{equation}
where $a_n$ are coefficients that can be determined order by order for parametrically small couplings. We write the $\sim$ sign and not an equality because
the series above, depending on the schemes involved, can be convergent or divergent asymptotic.
The two renormalization schemes can be qualitatively different, i.e. $g$ could correspond to a physical coupling and $g'$ to its minimally subtracted
${\rm MS}$ version in dimensional regularization, or they can be variants within the same family, say if $g$ is taken in $\overline{\rm MS}$.
In both cases \eqref{eq:RS1} applies, though the exact resummed change of scheme $g'=g'(g)$ could be completely different in the two cases,
in particular their analyticity properties. In renormalizable theories it is generally hard to go beyond \eqref{eq:RS1}, because the process of renormalization
occurs to all orders in perturbation theory. On the other hand, super-renormalizable theories require a finite number of subtractions and therefore provide a playground
for theories where we can hope to go beyond \eqref{eq:RS1} and find the exact form of the finite change of scheme $g'=g'(g)$.
We will do that in what follows for quartic $O(N)$-invariant scalar theories in $d=2$ and $d=3$ dimensions for a one-parameter family of renormalization schemes
within the same family  (like ${\rm MS}$ vs  $\overline{\rm MS}$).
The euclidean action of the theories read
\begin{equation}
  S = \int d^d x \Bigl[\frac12 (\partial_\mu \phi_i)^2+\frac 12 m_0^2 \phi_i^{2} + \lambda_0 (\phi_i^{2})^2  + \rho_0 \Bigr]\,, \quad i=1,\ldots, N\,.
  \label{eq:H0}
\end{equation}
As well-known, for $d<4$ only the mass term and the vacuum energy term require renormalization, the quartic coupling and the field $\phi$ being finite.
We consider dimensional regularization and define the renormalized mass and coupling in $d=2$ or $d=3$ as
\begin{equation}
  m^2_0 = m^2 + \delta m^2 \,, \quad
  \lambda_0 = \mu^\epsilon \lambda \,, \quad
  \rho_0 = \mu^{-\epsilon}(\rho+\delta\rho) \,,\quad  \quad
  d = n- \epsilon \,, \quad n=2,3\,.
  \label{eq:countDef}
\end{equation}
Note that $\lambda$ has mass dimension 2 and 1 in $d=2$ and $d=3$, respectively.
The introduction of an RG scale $\mu$ might confuse the reader. In fact, there are no large log's to be resummed
in perturbation theory and consequently no need to introduce a further mass scale $\mu$ in the problem. The natural choice would be
to simply set $\mu=m$ in \eqref{eq:countDef}. However, changing $\mu\rightarrow \mu \, e^{-\kappa/2}$, where $\kappa$ is an arbitrary real parameter, is equivalent to change the counterterms
$\delta m^2$ and $\delta \rho$ in \eqref{eq:countDef} and hence is a convenient way to introduce a simple one-parameter class of renormalization schemes.\footnote{In $4d$ a relation of this kind with $\kappa=\log(4\pi)-\gamma_{\rm E}$ links the ${\rm MS}$ and $\overline{\rm MS}$ schemes.}
That said, all the considerations below could be derived using e.g. cut-off regularization at fixed dimension, but at the price
of having more complicated expressions in $d=3$.

Let us assume that $m^2>0$, so that we are in the classically unbroken phase of the theory.
To all orders in perturbation theory the $\beta$ functions are easily determined since there are no contributions to $\beta_\lambda$
and only one to $\beta_{m^2}$ in both $d=2$ and $d=3$, given respectively by the first and second diagrams in fig.~\ref{fig:2pt_diags} of appendix~\ref{app:MVEd3}. One has \footnote{We do not report here the $\beta$ function for the vacuum energy, which for $d=3$ can be found in \eqref{eq:rhomu}, since it does not play any role in the analysis that follows. The vacuum energy will be neglected until section \ref{sec:Borel}.}
\begin{equation}
  \beta_\lambda = 0\,,  \quad \quad \beta_{m^2}  = 2 b_{d-1} \lambda^{d-1}\,, \quad d=2,3,
  \label{eq:RGmd}
\end{equation}
where
\begin{equation}
  b_1 = -\frac{N+2}{\pi} \,, \quad \quad b_2 = \frac{N+2}{\pi^2} \,.
\end{equation}

If we denote by $m^2$ the squared mass parameter in the original scheme, in the scheme where
$\mu\rightarrow \mu \, e^{-\kappa/2}$ we get a squared mass parameter $m'^{2}$ equal to
\begin{equation}
  m'^2(\mu) = m^2(\mu) + \lambda^{d-1} b_{d-1} \kappa\,.
  \label{eq:mSchems}
\end{equation}
Using the running of the mass term, this can be written more explicitly as
\begin{align}
  m^2 +\lambda^{d-1} b_{d-1} \log \frac{\mu^2}{m^2} & = m'^2 +\lambda^{d-1} b_{d-1} \log \frac{\mu^2}{m'^2} - \lambda^{d-1} b_{d-1} \kappa \,,
  \label{eq:m23d}
\end{align}
where $m^2(\mu^2=m^2) \equiv m^2$, $m'^2(\mu^2=m'^2) \equiv m'^2$.
The relation \eqref{eq:m23d} can be further rewritten as
\begin{equation}
  f_d(x)  = f_d(x^\prime) + \kappa \,,\label{eq:fggp}
\end{equation}
where
\begin{align}
  f_d(x) \equiv \log x + (-1)^d x \,, \quad \quad x\equiv \frac{1}{N+2} \Big(\frac{\pi}{g}\Big)^{d-1} \,,\quad \quad g \equiv \frac{\lambda}{m^{4-d}} \,.
  \label{eq:fdDef}
\end{align}
Note that $g$ is the dimensionless loopwise expansion parameter while the variable $x$ (in units of $\lambda$) is proportional to the squared mass term
in both $d=2$ and $d=3$ dimensions.
We can use \eqref{eq:fggp} to find an exact change of scheme $x^\prime=x^\prime(x)$.

Consider first the $d=2$ case. Solving for $x'$ we get for any $\kappa$ the unique solution
\begin{equation}
  x^\prime = W_0\Big(x e^{x-\kappa} \Big)\,,
  \label{eq:schemeWUn}
\end{equation}
where $W_0$ is the principal branch of the Lambert function $W$. This function will repeatedly appear in our considerations, so we refer the reader to appendix \ref{sec:Lambert} for its definition and a brief summary of some of its properties.
This solution agrees with the one obtained in perturbation theory by expanding $g^\prime$ for small $g$:
\begin{equation}
  g^\prime = g + \frac{(N+2)\kappa}\pi g^2 + \frac{(N+2)^2 \kappa (\kappa-1)}{\pi^2} g^3 + \ldots
  \label{eq:pertScheme}
\end{equation}
Instead of expanding $W$ for large values of its argument, that involves iterative logs, one can alternatively expand for small $\kappa$,
by noting that at order $n$ in perturbation theory the change of scheme
involves a polynomial of degree $n-1$ in $\kappa$. Defining $y = x- \kappa$, we are left with the expansion of $W(ye^y + \kappa e^y)$ for small $\kappa$.
The Taylor expansion around $\kappa=0$ can be performed using (\ref{eq:RRforL}) and the fact that by definition $W(ye^y) = y$. We get
\begin{equation}
  x^\prime = x-\kappa + \sum_{n=1}^\infty \frac{\kappa^n}{n!} \frac{p_n(x-\kappa)}{(1+x-\kappa)^{2n-1}}\,.
  \label{eq:xprimePer}
\end{equation}
Expanding this relation for large values of $x$ finally reproduces, to all orders in perturbation theory, the perturbative change of scheme
given in \eqref{eq:pertScheme}. We can use this formula to argue about the nature of the perturbative series associated to the change of scheme and
its radius of convergence. This can be easily determined in a ``double scaling" limit where\footnote{A scaling of this kind has been already considered in \cite{Sberveglieri:2019ccj}
  where the same class of 1-parameter family of renormalization schemes has been analyzed.}
\begin{equation}
  |\kappa| \rightarrow \infty \,, \quad x\rightarrow \infty\,, \quad \frac{x}{|\kappa|} \equiv \alpha = {\rm fixed}\,.
\end{equation}
For large $y$ we have
\begin{equation}
  p_n(y) = (-1)^{n-1} (n-1)! y^{n-1} + {\cal O}(y^{n-2}),
\end{equation}
from which we get
\begin{equation}
  \alpha^\prime \approx \alpha - \eta - \frac{1}{|\kappa|} \sum_{n=1}^\infty \frac{1}{n}(1-\eta \alpha)^{-n}\approx \alpha - \eta  +\frac{1}{|\kappa|}\log\Big(\frac{\alpha}{\alpha-\eta}\Big) \,,
  \label{eq:alphaprimeapprox}
\end{equation}
where $\eta \equiv {\rm sign} (\kappa)$.
The series giving rise to the log converges for $\alpha>2$ for $\kappa>0$ and $\alpha>0$ for $\kappa<0$.
Note that \eqref{eq:alphaprimeapprox} could be obtained more easily by using \eqref{eq:W0Largex}, though the above procedure simplifies the connection with the perturbative expansion of the change of scheme. The series is convergent also at finite $\kappa$, with a radius that depends on $\kappa$. The expansion of \eqref{eq:xprimePer} for large $x$ is of the form
\begin{equation}
  x^\prime = x-\kappa + \kappa \sum_{n=1}^\infty \frac{q_n(\kappa)}{x^n}\,,
  \label{eq:xprimePer2}
\end{equation}
where $q_n$ are polynomials of degree $n-1$ in $\kappa$.  The coefficients of the monomials entering $q_n$ alternate in sign and indicate that the convergence properties of the series are better when
$\kappa>0$. We have not determined the exact radius of convergence $R(\kappa)$ of the series \eqref{eq:xprimePer2} but we have checked that $R(\kappa) \sim 1/|\kappa|$, in line with the
analysis above.

\begin{figure}[t!]
  \includegraphics[width=0.5\textwidth]{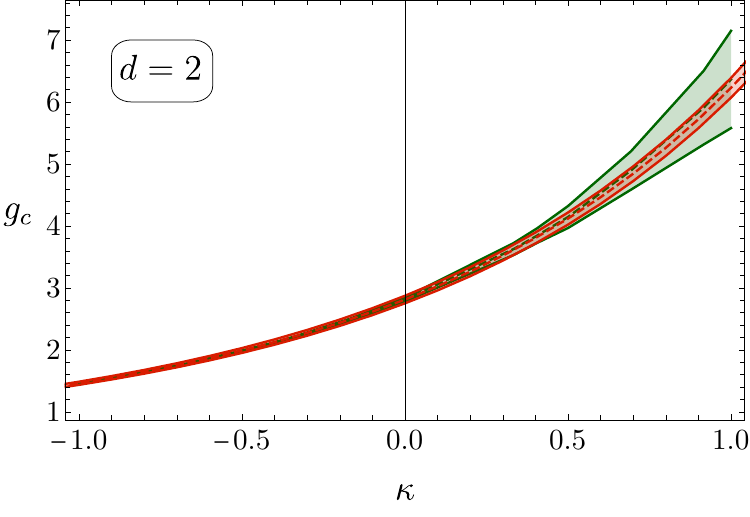}
  \centering
  \caption{In green the critical coupling $g_c$ of the $N=1$ $\phi^4$ theory in $d=2$, determined as $M(g_c) = 0$ using conformal mapping for several renormalization schemes parametrized by $\kappa$
    \cite{Sberveglieri:2019ccj}, in red the analytic curve for $g_c(\kappa)$ from (\ref{eq:gcKappa}) given as value of reference $g_c(\kappa=0)$.}
  \label{fig:gcritical_kappa_2d}
\end{figure}

It is well-known that the $N=1$ $\phi^4$ theory has a second-order phase transition at a critical value of the (inverse) coupling $x_c$ in the same universality class of the $d=2$ Ising model.\footnote{For $N=2$ vortices appear and the theory has a Berezinskii-Kosterlitz-Thouless transition \cite{Berezinsky:1970fr,Berezinsky:1972rfj,Kosterlitz:1973xp}. For $N\geq 3$ the theories are gapped and no transition occurs.
  See \cite{Gorbenko:2020xya} for a recent analysis of 2d $O(N)$ models for continuous values of $N$ between $-2$ and $2$.}  The dependence of $x_c$  on the renormalization scheme has been studied in \cite{Sberveglieri:2019ccj}, where it has numerically been found how $x_c$ depends on $\kappa$. Given $x_c(\kappa=0)\equiv x_c$, the exact relation \eqref{eq:schemeWUn} allows us to find the analytic form of  the dependence of the critical coupling
on the renormalization  scheme:
\begin{equation}
  x_c(\kappa) =  W_0(x_c e^{x_c-\kappa})\,.
  \label{eq:gcKappa}
\end{equation}
For $\kappa\rightarrow -\infty$, we have
\begin{equation}
  x_c(\kappa) \approx |\kappa| + x_c\,,
\end{equation}
and the fixed point coupling approaches the Gaussian one, while in the opposite limit $\kappa\rightarrow \infty$,
\begin{equation}
  x_c(\kappa) \approx x_c  e^{x_c} \,e^{-\kappa}
\end{equation}
the coupling goes to infinity exponentially in $\kappa$. We have verified that \eqref{eq:gcKappa} reproduces the results of fig.~3 of \cite{Sberveglieri:2019ccj}, see fig.~\ref{fig:gcritical_kappa_2d}.
As the renormalization scheme is varied, we always find a fixed point and hence the phase transition is ``visible" from the (classically) unbroken phase in $d=2$ for any choice of $\kappa$.

Let us now consider the more interesting case of $d=3$, where the apparently innocuous sign difference between $d=2$ and $d=3$ in \eqref{eq:fdDef} completely changes the
picture. Solving for $x'$, for any $\kappa$ we now get two solutions
\begin{equation}
  x^\prime_w = - W_{-1}\Big(- x e^{- x- \kappa} \Big)\,,\quad   x^{\prime}_s = - W_0\Big(- x e^{- x - \kappa} \Big)\,,
  \label{eq:schemeWBr}
\end{equation}
where $x^\prime_w>1$ and $x^\prime_s<1$, associated to the two different branches $W_{-1}$ and $W_{0}$ of the Lambert function, see appendix \ref{sec:Lambert}.
We label the two branches as weak ($w$) and strong ($s$)  branches. The solution that agrees with the one obtained in perturbation theory is obtained by expanding $W_{-1}$ for $x\rightarrow +\infty$, which corresponds to  $x e^{- x}\rightarrow 0^+$. Using \eqref{eq:Wm1x0}, we get
\begin{equation}
  x^\prime_w \approx x +\kappa + \ldots
\end{equation}
The perturbative change of scheme is obtained by proceeding as before and can be written in the form of \eqref{eq:xprimePer}, with the obvious replacement
$x\rightarrow- x$.
The other non-perturbative solution is obtained by expanding $W_{0}$ for $x\rightarrow +\infty$ and gives
\begin{equation}
  x^{\prime}_s  \approx  x e^{-x-\kappa}\,.
  \label{eq:strongx}
\end{equation}
Two solutions occur also for $\kappa=0$ and indicate that $O(N)$ vector models in $d=3$ admit two ``dual" descriptions in the classically unbroken phase.
They are related as follows:
\begin{equation}
  x_s = - W_0(-x_w e^{-x_w})\,, \quad {\rm or} \quad x_w = - W_{-1}(-x_s e^{-x_s})\,,
  \label{eq:strongxd}
\end{equation}
for $x_w>1$ and $x_s<1$.
In terms of mass scales the first relation in \eqref{eq:strongxd} gives
\begin{equation}
  \lim_{m\rightarrow \infty} m_s^2 \approx m^2 e^{-\frac{\pi^2 m^2}{(N+2)\lambda^2}}\,,
  \label{eq:massS3d}
\end{equation}
where $m_w^2 \equiv m^2$. Interestingly enough, \eqref{eq:massS3d} can be interpreted as the ``dynamically generated" RG invariant scale
\begin{equation}
  \Lambda_{{\rm RG}}^2 = \mu^2 e^{-\frac{\pi^2}{(N+2) g^2(\mu)}}\,,
  \label{eq:RGsclae3d}
\end{equation}
that arises from the $\beta$-function for $g^2$:
\begin{equation}
  \beta_{g^2} = -\frac{2(N+2)}{\pi^2} (g^2)^2\,.
\end{equation}
By taking $\mu = m$, $g(m) = \lambda/m$, we see that $\Lambda_{{\rm RG}}$ coincides with the weak coupling limit  \eqref{eq:massS3d} of $m_s$.
The strong and weak branch fuse at the self-dual point
\begin{equation}
  x_{{\rm SD}} = 1
  \quad \quad \Rightarrow \quad \quad
  g_{{\rm SD}} = \frac{\pi}{\sqrt{N+2}}\,,
  \quad \quad (d=3).
\end{equation}
In the large $N$ limit with $\lambda\rightarrow 0$, $N\rightarrow \infty$, and $\lambda N=$ fixed,
the two-loop term in \eqref{eq:m23d} drops out and correspondingly the function $f(x)$ trivializes. No self-duality survives in this large-$N$ limit.

A similar analysis can be done for $m^2<0$, namely in the classically broken phase, but
now the value of $N$ matters.  For $N>1$ in $d=2$ the Coleman-Mermin-Wagner theorem \cite{Coleman:1973ci,Mermin:1966fe} forbids the appearance of Goldstone bosons, so the theory is always non-perturbatively gapped and we cannot expect to be able to deduce strongly coupled effects by merely looking at perturbative counterterms.
This is in agreement with the fact that for $N>1$ in $d=2$ Borel summability is not guaranteed \cite{Serone:2018gjo}.
For $N>1$ in $d=3$ a continuous symmetry is spontaneously broken and massless Goldstone bosons appear.
The relation \eqref{eq:m23d}, based on the presence in the theory of a single $O(N)$-invariant mass scale, no longer holds and a more refined analysis is required (see also footnote \ref{footnote:Magruder}).
For simplicity, in what follows we focus on the case $N=1$, for which we expect that the analysis made above for $m^2>0$ also holds for $m^2<0$.\footnote{However, the situation simplifies at large $N$, where we can see the non-perturbatively generated mass gap from an analytic continuation in the coupling space, see subsection \ref{subsec:LargeNMassGap}.}
We denote the parameters in the broken phase with a tilde and continue to keep generic $N$ in the formulas, with
the understanding that $N=1$. The $\beta$-functions \eqref{eq:RGmd}  still apply, but we now have
\begin{equation}
  \mt^2(\mu^2=\mt^2) = - \frac 12 \mt^2\,,
  \label{eq:mtdef}
\end{equation}
where $-\mt^2/4$ is the renormalized mass term in the action, such that the particle excitation has squared mass $\mt^2>0$.
In the broken phase \eqref{eq:fggp} reads
\begin{equation}
  \widetilde f_d(\widetilde x)\  = \widetilde f_d(\widetilde x')+ \widetilde \kappa \,,\label{eq:fggptilde}
\end{equation}
where
\begin{align}
  \widetilde f_d(\xt) \equiv \log \xt - (-1)^d \xt \,, \quad \quad \xt \equiv \frac{1}{2(N+2)} \Big(\frac{\pi}{\gt}\Big)^{d-1} \,,\quad \quad \gt \equiv \frac{\lambda}{\mt^{4-d}} \,.
  \label{eq:fdDeftilde}
\end{align}
We see that, as far as the scheme dependence is concerned, the $d=2$ and $d=3$ theories in the broken phase behave respectively like
the $d=3$ and $d=2$ theories in the unbroken phase! The whole analysis made before applies with this replacement. In particular, we conclude
that the $d=2$ $N=1$ theory admits a self-duality in the broken phase.
The strong and weak branch fuse at the self-dual point
\begin{equation}
  \xt_{{\rm SD}} = 1
  \quad \quad \Rightarrow \quad \quad
  \gt_{{\rm SD}} = \frac{\pi}{2(N+2)}\,,
  \quad \quad (d=2).
\end{equation}

\section{Connection with Chang and Magruder Dualities}
\label{sec:Connec}

We have seen in section \ref{sec:RSDSD} how to perform an exact change of renormalization schemes within the same phase of the theory. However,
classically unbroken and broken phases are simply characterized by the sign of the squared mass term and since the latter is in fact divergent,
we should be able to push further our change of schemes (for $N=1$) and to relate one phase to another, passing through infinite coupling ($m^2=0$).
The relation \eqref{eq:mSchems} still applies and, in light of \eqref{eq:mtdef}, reads now
\begin{equation}
  \log (x/2) + (-1)^d  x  = \log \widetilde x - (-1)^d  \widetilde x+\kappa  \,,
  \label{eq:Changd2d3}
\end{equation}
in terms of the variables defined in \eqref{eq:fdDef} and \eqref{eq:fdDeftilde}.
The relation \eqref{eq:Changd2d3} states that a theory in the broken phase with negative squared mass term $-\mt^2/2$ is equivalent
to a theory in the unbroken phase with squared mass term $m^2$ (with the same $\lambda$) provided the two mass scales are related as in \eqref{eq:Changd2d3}.
The theories are ``dual" because they can be seen as the same theory where the mass term is renormalized differently.
For $\kappa=0$ in $d=2$,  the relation \eqref{eq:Changd2d3} coincides with Chang duality \cite{Chang:1976ek}, originally derived using a normal ordering prescription.
In $d=3$ relation \eqref{eq:Changd2d3} gives rise to a duality first discussed  by Magruder \cite{Magruder:1976px}.\footnote{Magruder actually wrote down a duality for arbitrary $N$
  by adding $O(N)$ group theoretical factors to the $N=1$ case, as if the symmetry would be linearly realized,  see (3.17) of \cite{Magruder:1976px}.
  For instance, the  term proportional to $\Lambda-\mu$ on the right hand side of  the counterterm (3.16) in \cite{Magruder:1976px} would naturally arise if all particles in the one-loop
  tadpole-like diagram responsible for the linear divergence had mass $\mu^2$. Massless particles would induce IR divergences in the sunset diagram contribution,
  proportional to $\log(\mu/\Lambda)$ in (3.16). Due to the derivative interactions of Goldstone bosons, we expect that IR divergences cancel, but in a non-trivial way
  in a linear parametrization in terms of the field-components of $\phi_i$. A duality might still hold for $N>1$ but establishing it requires to
  understand how to map operators in a theory from a phase to another, where a global symmetry is linearly or non-linearly realized, respectively. \label{footnote:Magruder}}
The original derivation of \cite{Magruder:1976px} made use of cut-off regularization and a different renormalization scheme (without the need of introducing a RG scale $\mu$), where
an extra term proportional to $\sqrt{x}$ appeared on both sides of \eqref{eq:Changd2d3} due to a divergence induced by the one-loop tadpole-like diagram.
The presence of such term hinders an analytic solution of the duality relation and it obscures the close analogy between the $d=2$ and the $d=3$ cases.
This divergence depends on the renormalization scheme and is set to zero in minimal subtraction schemes based on dimensional regularization.
Note that no duality occurs for non-integer $d$, since the log terms in \eqref{eq:Changd2d3} can only appear for integer dimensions.

\begin{figure}[t!]
  \includegraphics[width=0.8\textwidth]{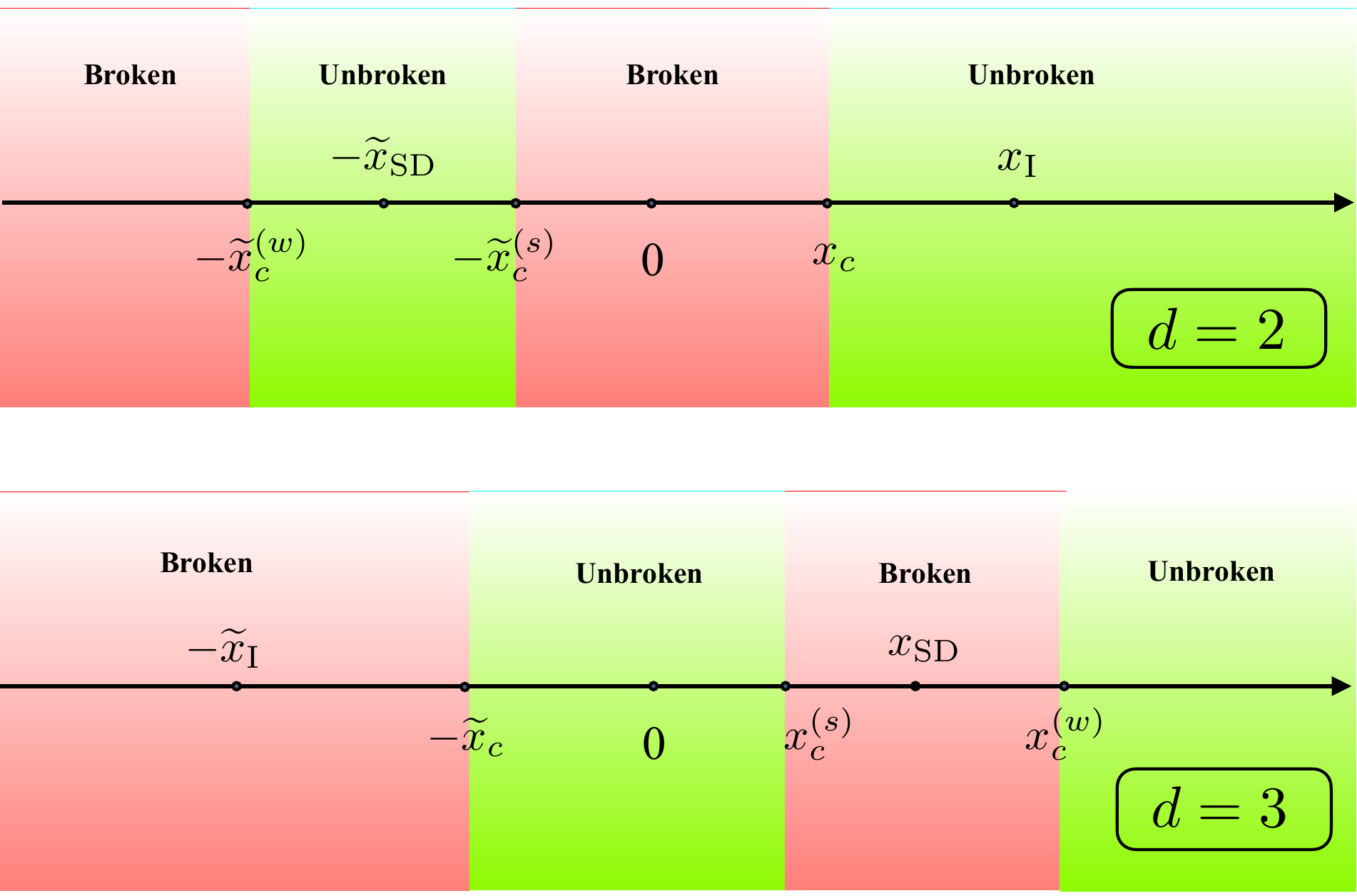}
  \centering
  \caption{Phase structure of the $N=1$ $\phi^4$ theory according to the Chang-Magruder dualities in $d=2$ and $d=3$, in schemes where respectively $x_c < x_{{\rm I}}$ and $\xt_c < \xt_{{\rm I}}$, as a function of the parameter $x$ (proportional to the squared mass).   }
  \label{fig:ChangRS}
\end{figure}
For $\kappa=0$ and at fixed $x$ ($\widetilde x$), the solutions in $\widetilde x$ ($x)$ of (\ref{eq:Changd2d3}) are
\begin{align}
  \widetilde x_w & = - W_{-1}(- \omega x e^x )\,, \quad \quad  \widetilde x_s = - W_0(- \omega x e^x )\,, \quad \quad   x   = W_0\Big(\frac{\widetilde x}{\omega} e^{-\widetilde x} \Big) \quad \quad \;\;(d=2) \,, \label{eq:Ch2d} \\
  \widetilde x   & = W_0(x \omega e^{-x})\,, \quad \quad \quad   x_s   = -W_0\Big(-\frac{\widetilde x}{\omega} e^{\widetilde x} \Big) \,,
  \quad \quad   x_w   = -W_{-1}\Big(-\frac{\widetilde x}{\omega} e^{\widetilde x} \Big)\quad (d=3) \,,
  \label{eq:Ch3d}
\end{align}
with $\omega=1/2$. Note that \eqref{eq:Ch2d} and \eqref{eq:Ch3d} are related by the map
\begin{equation}
  x\leftrightarrow \xt \,, \quad \quad  \omega \leftrightarrow \frac{1}{\omega}\,,
\end{equation}
which is again a manifestation of the interplay between unbroken and broken phases in $d=2$ and $d=3$.
In $d=2$, at fixed $x$, the two solutions in (\ref{eq:Ch2d}) are real for
$xe^x/2 < 1/e$, i.e. for  (setting $N=1$)
\begin{equation}
  g \geq g_\mathrm{I} \equiv \frac{\pi}{3W_0(2/e)} \approx 2.26\,,
  \quad (d=2)\,.
\end{equation}
In $d=3$, at fixed $\widetilde x$, the two solutions in (\ref{eq:Ch3d}) are real for $2 \xt e^{\xt } < 1/e$, i.e. for   (setting $N=1$)
\begin{equation}
  \gt \geq \gt_\mathrm{I} \equiv \Big(\frac{\pi^2}{6W_0(1/(2e))} \Big)^{1/2}\approx 3.23\,,
  \quad (d=3)\,.
\end{equation}
Depending on the value of the coupling, the theories admit one or three equivalent descriptions. We summarize the phase structure in fig.~\ref{fig:ChangRS}. In $d=2$  the theory admits only one description in the classically unbroken phase for $x > x_\mathrm{I}$, where $x_\mathrm{I}$ is the map of the self-dual point $\widetilde x_\mathrm{SD}$ by means of \eqref{eq:Changd2d3}.
The region $0 < x < x_\mathrm{I}$ can instead be mapped to $0 < \widetilde x < \widetilde x_\mathrm{SD}$ and $\widetilde x > \widetilde x_\mathrm{SD}$, so three descriptions are possible,
one in the classically unbroken and two in the classically broken phases. Within our class of schemes the position of the self-dual coupling is invariant while the positions of the critical couplings, denoted by $x_c$, $\xt_c^{(w)}$ and $\xt_c^{(s)}$ with obvious notation, depend on the renormalization schemes and are related, as discussed in the next section. In the schemes where $x_c > x_{{\rm I}}$ in $d=2$, the unbroken region around $\xt_{{\rm SD}}$ disappears and the phase transition is accessible only from the unbroken phase. In $d=3$ the structure is the same after the substitutions $x\leftrightarrow \xt$ and inverting the role of broken and unbroken phases.

\section{Fixed Points Annihilation and Analyticity Domain}

\label{sec:FPAAD}

It is well-known that $d=3$ $O(N)$ quartic models undergo
a second-order phase transition for any value of $N$. So, how could one trust the existence of a strong-weak duality in these theories based on perturbative treatments around the (classically) unbroken phase? In particular, where is the broken phase? We will now address these questions.
\begin{figure}[t!]
  \includegraphics[width=0.45\textwidth]{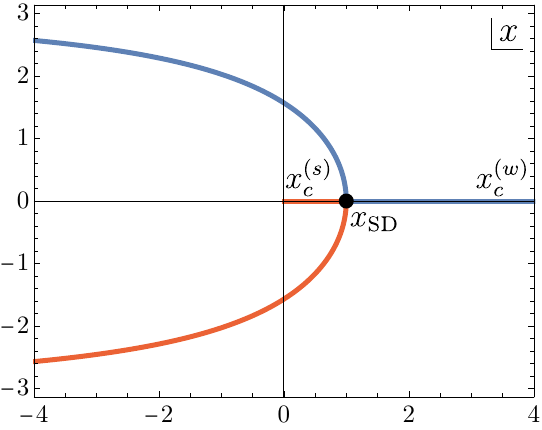}
  \centering
  \caption{Positions in the complex $x$ plane of the critical values of the weak ($x_c^{(w)}$ blue line) and strong ($x_c^{(s)}$ red line) branches as the renormalization scheme $\kappa$ is varied.
  The black dot corresponds to the self-dual point $x_\mathrm{SD} = 1$
  where the critical points merge.
  }	\label{fig:xplane}
\end{figure}

Suppose that in a given renormalization scheme the $d=3$ $O(N)$ models have a phase transition for $o (1)$ real values of $x_c^{(w)}$ and $x_c^{(s)}$ in the weak and strong branches, respectively. The existence of such schemes will be proved in section \ref{sec:Borel}.
Using \eqref{eq:schemeWBr} we can determine how the fixed points move when we change renormalization scheme:
\begin{align}
  x_c^{(w)}(\kappa) & =  - W_{-1}\Big(-x_c^{(w)} e^{-x_c^{(w)}- \kappa}\Big)  \,, \nn \\
  x_c^{(s)}(\kappa) & = - W_{0}\Big(-x_c^{(s)} e^{-x_c^{(s)}- \kappa}\Big)  \,,
  \label{eq:gcFunkappa}
\end{align}
where $x_c^{(w)}\equiv x_c^{(w)}(\kappa=0)$, $x_c^{(s)}\equiv x_c^{(s)}(\kappa=0)$.\footnote{Note that the parameter $\kappa$ in \eqref{eq:gcFunkappa} is shifted by a constant with respect to the $\kappa$
  defined in section \ref{sec:Borel}.}
For $ \kappa>0$, as $ \kappa$ increases, $x_c^{(w)}(\kappa)$ and $x_c^{(s)}(\kappa)$ respectively increases and decreases, moving far apart. On the other hand,
for $ \kappa<0$,  as $| \kappa|$ increases $x_c^{(w)}(\kappa)$ and $x_c^{(s)}(\kappa)$ respectively decreases and increases, approaching each other, until they merge
when the argument of the two branches of the Lambert function equal $-1/e$, namely at the self-dual point
\begin{equation}
  x_c^{(w)}(\kappa_*)  = x_c^{(s)}(\kappa_*) =x_{{\rm SD}}  =1\,, \quad  \kappa_* = 1+\log(x_c e^{-x_c})\,.
\end{equation}
For $\kappa$ slightly smaller than $\kappa_*$,  $x_c^{(w)}$ and $x_c^{(s)}$ move in the imaginary axis in a complex conjugate pair.
As $\kappa$ decreases they move backwards in an approximate parabolic trajectory and then they move towards $|x|\rightarrow \infty$ in parallel along the negative real axis with
${\rm Im} \,x_c^{(w)} \rightarrow \pi$, ${\rm Im} \, x_c^{(s)} \rightarrow -\pi$, see fig.~\ref{fig:xplane}.
More precisely, for $\kappa\rightarrow -\infty$, we have
\begin{equation}
  x_c^{(w)} (\kappa) \approx- |\kappa| + i \pi  \,,\qquad   x_c^{(s)} (\kappa) \approx- |\kappa| - i \pi  \,,
\end{equation}
and both critical couplings approach the free theory, while for $\kappa\rightarrow \infty$ we have
\begin{equation}
  x_c^{(w)} (\kappa) \approx \kappa + x_c^{(w)}\,, \qquad
  x_c^{(s)} (\kappa) \approx x_c^{(s)}  e^{- x_c^{(s)}} \,e^{-\kappa}\,.
\end{equation}

This implies that for $ \kappa <  \kappa_*$ {\it the phase transition  cannot be seen in $d=3$ $O(N)$ theories  when
    starting from the classically unbroken phase, with real values of the coupling}.
In these schemes we can then hope to have access to the strong-weak duality starting from perturbative considerations, without encountering non-analyticities associated
with phase transitions. Note that this is independent of the Magruder duality and hence apply for any $N$.

A few studies of the 2d and 3d $\phi^4$ theories have been performed away from criticality, and there was no consensus on the appearance of first/second-order phase transitions when $m^2>0$.
Studies using the Gaussian effective potential were either inconclusive on the appearance of a phase transition \cite{Cea:1996pe} or found a phase transition that could be first or second-order \cite{Stancu:1990yf}.
Ref.\cite{Windoloski:2000yb} studied the $\phi^4$ theory at finite volume using Monte Carlo and finite states truncations, and found no phase transition for $m^2>0$.
We see that this problem was in fact a red herring, since the appearance of the phase transition (more precisely a gapless phase for real values of the coupling) for $m^2>0$ is a renormalization scheme-dependent question.

\begin{figure}[t!]
  \includegraphics[width=0.49\textwidth]{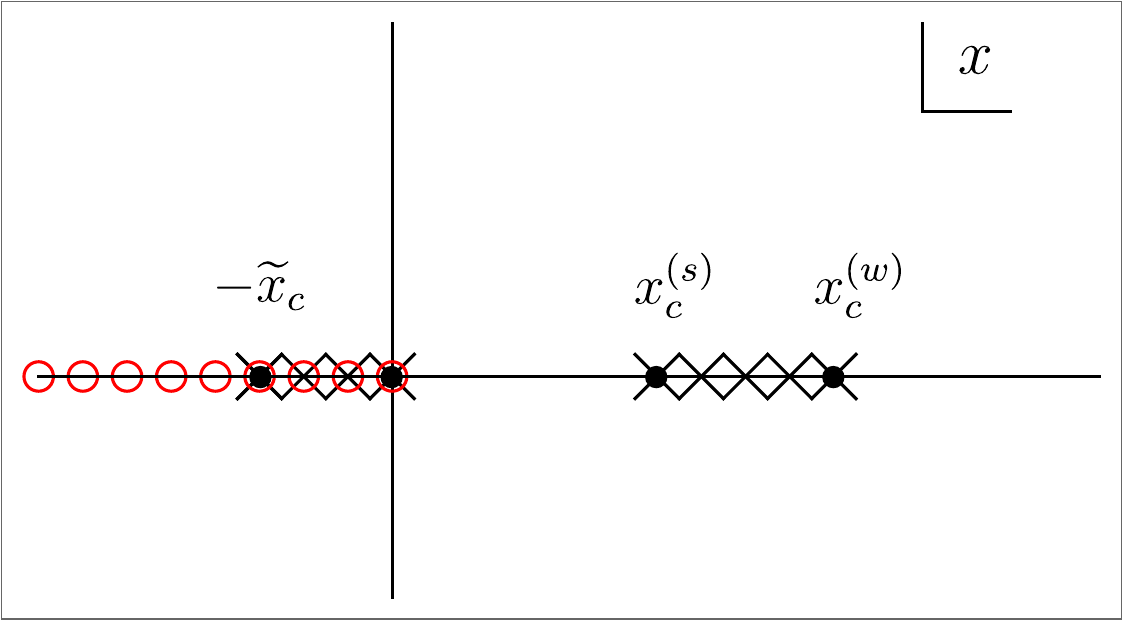}\;\;\;
  \includegraphics[width=0.48\textwidth]{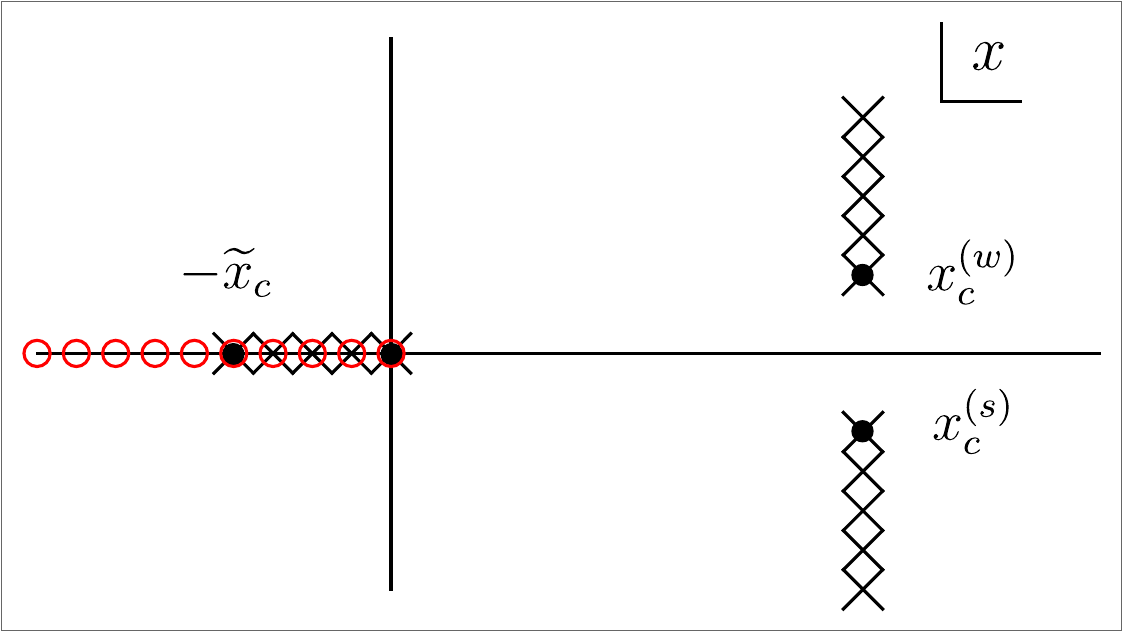}
  \centering
  \caption{Conjectured minimal singularity structure of observables as analytic functions of the coupling $x$ (proportional to $m^2$) for $O(N)$ vector theories in $d=3$ for $\kappa>\kappa_*$  (left) and $\kappa< \kappa_*$ (right).}
  \label{fig:Xana}
\end{figure}

We can speculate about the analyticity properties of generic observables $F(x)$ as analytic functions of $x$, see fig.~\ref{fig:Xana}.\footnote{A relevant class of observables are Schwinger $n$-point functions smeared with Schwarzian test functions.} We expect that $F(x)$ should have a branch-cut singularity at infinity, which corresponds to the usual branch-cut
associated to perturbative asymptotic expansions around free theories. Self-duality implies that the origin should also be a singular branch-point. In the assumption of maximal analyticity, 
the branch-cut at infinity and the one at the origin are continuously connected. This branch-cut is depicted by red circles in fig.~\ref{fig:Xana}. In addition to that, we expect
further branch-cut singularities in correspondence of the critical values $x_c^{(w)}$ and $x_c^{(s)}$, either on the real line or in the complex plane, depending on the choice of renormalization scheme.
For $N>1$ we do not really know the analytic structure in the classically broken phase (${\rm Re}\ x<0$). Assuming again maximal analyticity, we might have a single critical value on the real line at $-\widetilde x_c$ for any $\kappa$, as expected in the $N=1$ case. The further branch-cuts associated to $x_c^{(w)}$, $x_c^{(s)}$ and $-\widetilde x_c$ are depicted by black crossed lines in fig.~\ref{fig:Xana}. 
These are the minimal singularities that we expect in the complex $x$ plane, but of course others could be present.
It would be extremely interesting to understand if the analyticity properties of observables, together with perturbative data and the self-duality condition $F(x_w) = F(x_s)$ might allow for an exact solution for the $O(N)$ models.

\subsection{Large $N$ Non-Perturbative Mass Gap in $d=2$}
\label{subsec:LargeNMassGap}

It is well-known that the appearance of a non-perturbative mass gap can be derived at leading order in a $1/N$ large $N$ expansion in $d=2$ $O(N)$ vector models \cite{Coleman:1974jh}.
We will see here how such a mass gap can be interpreted to arise from an analytic continuation of the squared mass from positive to negative values.
By introducing a Hubbard-Stratonovich (HS) field $\sigma(x)$ we can rewrite $S$ as in~\eqref{eq:Ssigma}. Neglecting the vacuum energy and the counterterm $\hat \delta m^2$, sub-leading in $o(N^{-1})$, we have
\begin{equation}
  \hat S =  \int d^d x \Bigl[\frac12 (\partial_\mu \phi_i)^2+\frac 12 m^2 \phi^{2}_i -\frac 12 \sigma^2+\frac 12 \hat f \sigma  \phi^{2}_i+ \sigma \delta_T \Bigr]\,.
  \label{eq:SsigmaFiniteN}
\end{equation}
Note that $m^2$ in \eqref{eq:SsigmaFiniteN} can be positive or negative.
If we integrate out the scalar fields $\phi_i$ we get an effective potential for $\sigma$. Its extremum is given in the \MSb scheme by
\begin{equation}
  \sigma = - \frac{m^2}{\hat f} + \frac{N \hat f}{8 \pi} W(M^2) \,,
  \label{eq:Lam}
\end{equation}
where
\begin{equation}
  M^2 = \frac{\pi \mu^2}{\hat \lambda} e^{\frac{\pi m^2(\mu)}{\hat \lambda}}
\end{equation}
and $\hat \lambda$ is fixed in the large $N$ limit, see \eqref{eq:largeNlimit}. Note that $M$ is an RG-invariant scale with respect to the large $N$ limit of the $\beta$-function in \eqref{eq:RGmd}.
In particular, we can set $\mu^2=|m|^2$. The value of $m^2(\mu^2=|m|^2)$ corresponds by definition to the classical mass term in the action.
In the classically unbroken case, we have $m^2>0$ and \eqref{eq:Lam} boils down to $\sigma=0$, since $W(xe^x)=x$ by definition and the two terms
in \eqref{eq:Lam} cancels each other ($8\hat\lambda = N \hat f^2$). As expected, the HS field gets no VEV in the unbroken case and the gap in the theory is determined by the classical mass term $m$.
On the other hand, in the classically broken phase $m^2<0$ and the Lambert function does not ``trivialize". Correspondingly the HS field gets a VEV,  the classical $m^2$ term
in \eqref{eq:SsigmaFiniteN} is cancelled by the first term in \eqref{eq:Lam} and we are left with a positive non-perturbative mass term equal to
\begin{equation}
  m_{{\rm np}}^2 = \frac{\hat \lambda}{\pi} W(M^2)\,.
\end{equation}
In the parametric weakly coupled limit $\hat \lambda/|m^2|\rightarrow 0$, we have
\begin{equation}
  m_{{\rm np}}^2 \approx  |m^2| \; e^{-\frac{\pi |m^2|}{\hat\lambda}} \,.
  \label{mnpNP}
\end{equation}
Both the perturbative and non-perturbative mass gaps arise from \eqref{eq:Lam}. We can then also interpret the non-perturbative mass gap as the analytic continuation
of the perturbative one from $m^2>0$ to $m^2<0$, passing through infinite coupling.\footnote{ A mass gap seen as analytic continuation past infinity in the large $N$ limit of non-linear $O(N)$ sigma models has been suggested in \cite{Yamazaki:2019arj}.} Interestingly enough, the non-perturbative scale \eqref{mnpNP} can also be deduced from IR renormalons that would appear in a perturbative expansion around the
``naive" vacuum $\sigma=0$ \cite{Marino:2019fvu}.

\section{Numerical Results in $d=3$ $O(N)$ Models}
\label{sec:Borel}

We report in this section the results obtained by resumming the perturbative series for the vacuum energy and the mass gap defined as
\begin{equation}
  \Lambda\equiv \Gamma^{(0)}\,, \quad \quad M^2 \equiv \Gamma^{(2)}(p=0)\,,
\end{equation}
as a function of the coupling $g$ in 3d $O(N)$ vector models. We confirm the theoretical expectations made in the previous sections. In particular, we  provide evidence for the self-duality of these models and determine how the critical coupling $g_c$ depends on the renormalization scheme.  We used, as in our previous works, two independent methods for the resummation: conformal mapping and reconstruction of the Borel function via Pad\'e approximants (in the following denoted for short conformal-Borel and Pad\'e-Borel respectively). We do not report the details of the numerical implementation, which can be found in \cite{Serone:2018gjo}.\footnote{Pad\'e approximants with poles on the positive real axis of the Borel variable were excluded in \cite{Serone:2018gjo}. These are now included taking the Cauchy principal value and adding  to the error estimate the residue at the pole.}
The parameters needed to perform the conformal mapping in 3d $O(N)$ models are well-known and can be found e.g. in \cite{Brezin:1992sq}.
In all our results we find agreement between conformal-Borel and Pad\'e-Borel methods, typically with slightly smaller uncertainties in the first one, and a consistent
convergence of the results as the number of loops used in the resummation is increased.
For this reason, in order to avoid clutter in the figures, we have decided to only plot quantities computed using conformal-Borel to the maximum available order.

\subsection{Perturbative Coefficients up to $g^8$}
\label{subsec:coeff}

We have computed the perturbative expansion of the zero-point function and the two-point function at zero external momentum up to order $g^8$. The computation has been performed numerically in momentum space using various simplifications introduced in \cite{Baker:1976ff,Nickel:1978ds,Guida:2005bc}. In the following we summarize the principal aspects of the computation.

\paragraph*{Choice of the scheme.} Since we compute loop integrals numerically,  a direct use of dimensional regularization  is unfeasible. It is instead convenient to regularize divergences
without introducing a regulator, subtracting to integrands of Feynman diagrams their values at a given fixed momentum, as proposed long ago by Zimmermann \cite{Zimmermann:1969jj}.
In this intermediate scheme (labeled with the subscript $I$) the mass counterterm $\delta m^2_I$ not only removes the divergence coming from the sunset-diagram (here chosen in such a way that the sunset diagram is regularized to be exactly zero at $p=0$) but it cancels also the one-loop tadpole-like diagram:
\begin{equation}
  \delta m^2_I = - \left( ~\raisebox{-0.4cm}{\includegraphics[height=1cm]{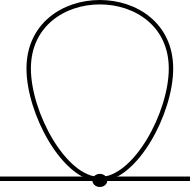}} + \raisebox{-0.4cm}{\includegraphics[height=1cm]{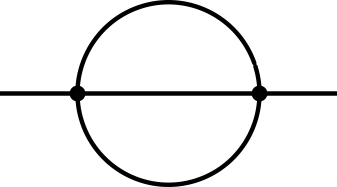}}{}^{\hspace*{-8pt} p=0} ~\right) \,.
  \label{eq:dm2I}
\end{equation}
Renormalization of higher order diagrams is then trivially implemented by substituting every tadpole and sunset subdiagram by its regularized counterpart:
\begin{align}
  \raisebox{-0.4cm}{\includegraphics[height=1cm]{Figs/2pt_o1.png}}_{\,\mathrm{reg}}                                    & = 0 \,,
  \\
  \raisebox{-0.4cm}{\includegraphics[height=1cm]{Figs/2pt_o2_a.png}}_{\hspace*{-10pt}\mathrm{reg}}{}^{\hspace*{-7pt}p} & =
  -\lambda^2\frac{N+2}{\pi^2}  \left[ 2 -\log \left(1+\frac{p^2}{9 m^2_I}\right)-\frac{6 m_I}{|p|} \arctan\left(\frac{|p|}{3m_I}\right) \right]\,.
  \label{eq:sunset_subd}
\end{align}
All the diagrams involving tadpoles are set to zero, greatly reducing the number of integrals to compute.
The vacuum energy counterterm in this scheme is chosen such that all contributions up to $o(g^3_I)$ vanish.

\paragraph*{Simplification of the integrands and numerical computation.} In order to improve the efficiency of the numerical integration we performed some analytical simplifications on the integrands that allowed us to greatly reduce the cost of the integrals for every diagram. In particular, we substituted the one-loop subdiagrams within the main diagram with two, three, and four external legs with their analytical expression \cite{Baker:1976ff,Nickel:1978ds}. Other simple subdiagrams that can be substituted are
\begin{align}
  \raisebox{-0.4cm}{\includegraphics[height=1.2cm]{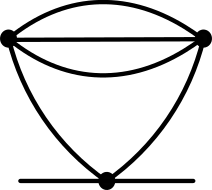}}^{\hspace*{-4pt}\mathrm{reg}}  =
  - \lambda^3 \frac{(N+2)^2}{m_I \pi^3} \log\Big(\frac43 \Big)\,,
  \qquad
  \raisebox{-0.5cm}{\includegraphics[height=1.2cm]{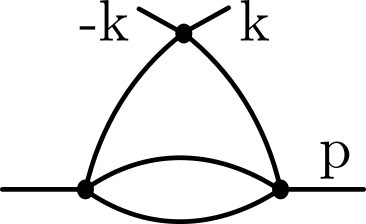}}                               =
  48 \lambda^3 \frac{5N+22}{\pi^2 m_I |p|} \arctan\left(\frac{|p|}{3m_I}\right) \,,
\end{align}
where in the first diagram we used the regularized sunset subdiagram of \eqref{eq:sunset_subd} and in the second diagram there is zero net momentum flow from the top vertex. Furthermore, by switching to spherical coordinates some of the integrals over angular variables can be performed analytically.
We have then numerically integrated each diagram using the Monte Carlo VEGAS algorithm \cite{Lepage:1977sw} from the python module \texttt{vegas} and later combined all the results with their corresponding $O(N)$ symmetry factors. As a sanity check, we compared the large $N$ limit of the perturbative expressions for $\Lambda$ and $M^2$ so obtained with those directly computed using large $N$ techniques
and found total agreement within the accuracy of the numerical evaluation of Feynman diagrams. We report in appendix \ref{sec:LL4D} the computation of $\Lambda$ and $M^2$ at the first non-trivial order
in the large $N$ limit. As a further check, we have computed the series of $d\Gamma^{(2)}/dp^2(p=0)$ and $\Gamma^{(4)}(p=0)$ up to order $g^8_I$. In this way, as explained in section 6.1 of \cite{Serone:2018gjo},  we can determine the series expansion of the $\beta$-function and of the critical exponent $\eta$ in the physical scheme of \cite{Parisi:1993sp} and have verified that they match with those appearing in the literature, known up to order $g^7$ and $g^6$ respectively \cite{Antonenko:1998es}. We hope to come back to the analysis of the critical theory in the scheme of  \cite{Parisi:1993sp}
in a future work.

\paragraph*{Mapping to the \MSb scheme.}

As a last step we have switched to the \MSb scheme by perturbatively reexpanding  $m^2_I(m^2)$ in powers of $\lambda$. The matching of the schemes is obtained by imposing the relation
\begin{equation}
  m_I^2 + \delta m_I^2 = m^2 + \delta m^2 \,,
\end{equation}
with $m^2$ and $\delta m^2$ in the \MSb scheme and we write $\delta m^2_I = -\Sigma_1 - \Sigma_{2a}(0)$. The explicit expressions for  $\Sigma_1$, $\Sigma_{2a}$ and $\delta m^2$ can be found in the appendix~\ref{app:MVEd3}. We get
\begin{equation}
  m_I^2
  =
  m^2 - \lambda m_I \frac{N+2}{\pi} + \lambda^2 \frac{N+2}{\pi^2} \left(\log \frac{9 m_I^2}{m^2} -1 \right) \,.
\end{equation}
By iteratively substituting $m_I$ in the right-hand side we then find the sought expansion. The first three orders are
\begin{equation}
  m_I^2 = m^2
  \left[
    1
    - g \frac{N+2}{\pi }
    + g^2 \frac{ (N+2) (N+4\log 3)}{2 \pi ^2}
    - g^3 \frac{ (N+2)^2 (N+6+8 \log 3)}{8 \pi ^3} + o\left(g^4\right)
    \right] \,.
\end{equation}
The  vacuum energy is divergent up to order $g^3$ and needs to be regularized by a vacuum energy counterterm $\delta \rho$. The computation of diagrams up to $o(g^3)$ in the \MSb scheme is presented in appendix~\ref{app:MVEd3}.  The final Taylor expansion up to order $g^8$ of  both $\Lambda$ and $M^2$ in the \MSb scheme is reported in appendix \ref{app:Coeff3d}.
We can now derive the series for $\Lambda$ and $M^2$ for the whole one-parameter class of renormalization schemes presented in section \ref{sec:RSDSD}. We identify $\kappa=0$ with the \MSb scheme above.
Starting from this, it is straightforward to compute the perturbative series in a generic scheme parametrized by $\kappa$ by using the expansion of \eqref{eq:schemeWBr}.
We refrain to write the whole lengthy series for $\Lambda$ and $M^2$ as a function of $N$ and $\kappa$. For illustration, we just report below the terms up to $o(g^2)$ 
in both series:
\begin{align}
  \frac{M^2}{m^2}          & = \ 1 - g \frac{N+2}{\pi } +g^2 \frac{(N+2) (N+4\log 3-2\kappa)}{2 \pi ^2}+\dots\,,                             \label{eq:M2Lm_kapdep}  \\
  \frac{\Lambda-\rho}{m^3} & = -\frac{N}{12\pi} + g \frac{N(N+2)}{16\pi^2} - g^2  \frac{N(N+2)}{8\pi^3} \left(\frac{N+2}{4} -3 + 4\log 2-\kappa\right)+\dots \,. \nn
\end{align}
For simplicity of notation the dependence on $\kappa$ of the parameters $m^2$, $g$ and $\rho$ has been left implicit in \eqref{eq:M2Lm_kapdep}.
Note that the series above could equivalently be interpreted as the series in the \MSb scheme with $\kappa=0$, but with parameters $m^2$ and $\rho$ evaluated at the scale $\kappa = \log (\mu^2/m^2)$. In this way, a sanity check of the validity of the change of scheme is obtained by demanding that both $M^2$ and $\Lambda$ satisfy the Callan-Symanzik equations
\begin{equation}
  \begin{split}
    \big( \mu \partial_\mu+ \beta_{m^2} \partial_{m^2} \big) M^2 & = 0 \,,  \\
    \big( \mu \partial_\mu+ \beta_{m^2} \partial_{m^2} +\beta_{\rho} \partial_{\rho} \big) \Lambda & = 0 \,,
  \end{split}
\end{equation}
with $\beta_{m^2}$ and $\beta_\rho$ given by \eqref{eq:RGmd} and \eqref{eq:rhomu}, respectively.
We always normalize the vacuum energy as $\rho(\kappa=0)=\rho(m)=0$. This implies that in computing $\Lambda$ in a scheme with $\kappa\neq 0$ the parameter $\rho(\kappa)$ is non-vanishing and should be taken into account.

\begin{figure}[t!]
  \includegraphics[width=0.487\textwidth]{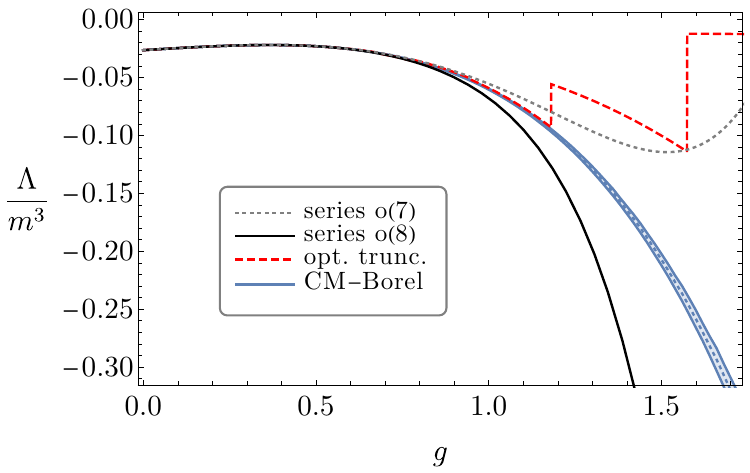}\quad
  \includegraphics[width=0.487\textwidth]{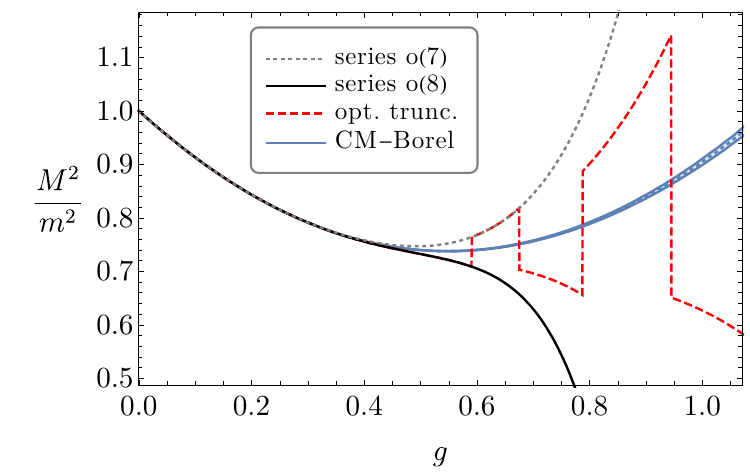}
  \centering
  \caption{The vacuum energy $\Lambda$ and  the mass gap $M^2$ as a function of the coupling constant $g$
    obtained by ordinary perturbation theory up to $g^7$ and $g^8$ (dotted grey and black lines), optimal truncation (red dotted line) and conformal-Borel resummation (blue line).}
  \label{fig:Lambda_M2_PT}
\end{figure}

\subsection{Self-Duality}
\label{sec:self-dual}

We report here the results obtained by numerical Borel resummation of the perturbative series for $\Lambda$ and $M^2$ for different values of $N$ and provide evidence for the self-duality of 3d $O(N)$ vector models. We start by showing the need of resumming the perturbative series in the region of couplings of interest. To this purpose,
we compare in fig.~\ref{fig:Lambda_M2_PT} $\Lambda$ and $M^2$ as a function of the coupling $g$ computed using the perturbative seven and eight loop results, optimal truncation, and Borel resummations. We take $N=1$ and choose the renormalization scheme $\kappa =0$, where $M^2$ does not vanish for real values of $g$. A similar analysis applies for other values of $N$.
In both figures it is clear that perturbation theory breaks down before $g_\mathrm{SD}=\pi/\sqrt{3}$ at a value of $g\approx 1$ for $\Lambda$ and $g\approx 0.6$ for $M^2$, and we observe that these values slightly decrease while increasing $N$. Therefore resummation techniques are required in order to study the self-duality.

\begin{figure}[t!]
  \includegraphics[width=0.487\textwidth]{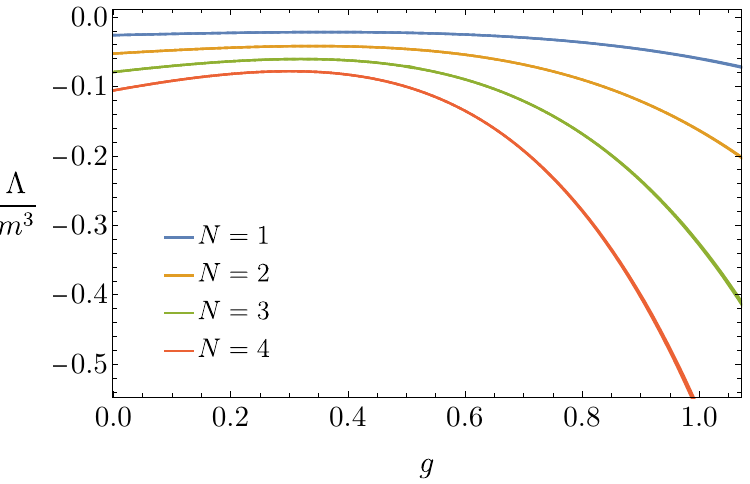}\quad
  \includegraphics[width=0.487\textwidth]{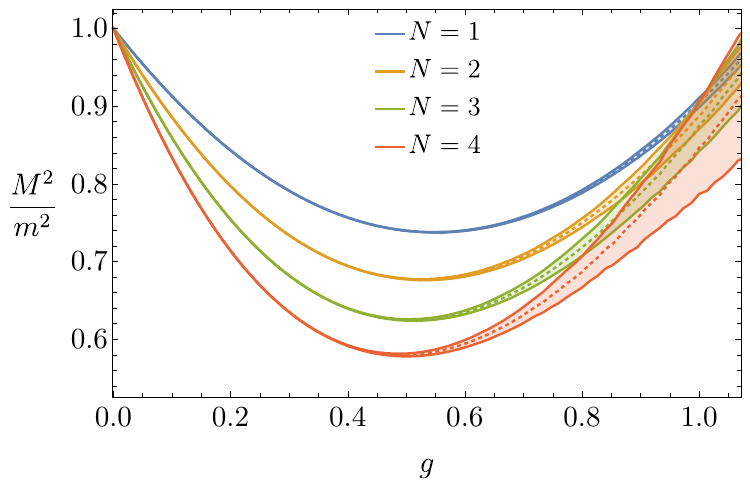}
  \centering
  \caption{The vacuum energy $\Lambda$ and the mass gap $M^2$ as a function of the coupling constant $g$ for different values of $N$ in the scheme $\kappa=0$.
    The results shown correspond to conformal-Borel resummation.}
  \label{fig:Lambda_M2}
\end{figure}

As discussed in the previous sections, for $\kappa< \kappa_*$ the phase transition is expected to be not visible from the unbroken phase.
We show in fig.~\ref{fig:Lambda_M2} $\Lambda$ and $M^2$ as a function of the coupling $g$ at $\kappa=0$ computed for different values of $N$ and using conformal-Borel resummation.
The right panel in fig.~\ref{fig:Lambda_M2} clearly shows that $M^2(g)$ is always positive, with the curve $M^2(g)$ developing a minimum around $g=0.6$ and then continuing to increase for larger values of $g$, as shown in fig.~\ref{fig:M2_kappa} for $N=1$; this confirms the absence of a gapless phase. We can determine $\kappa_*(N)$ by computing the critical coupling for a value of $\kappa$ where the transition occurs and then
use the map to determine the values of $\kappa_*(N)$ where the two critical points merge. Taking as reference value $\kappa=5$, we get for the first values of $N$
$\kappa_*(1)=3.5(2)$, $\kappa_*(2)=3.3(2)$, $\kappa_*(3)=3.2(2)$ and $\kappa_*(4)=3.1(3)$.

Let us now focus on the region $\kappa<\kappa_*$, where $M^2$ vanishes for complex values of the coupling, and discuss the self-duality.
First of all let us explain why we can probe self-duality using resummations of the perturbative series.
The complex points where $M^2$ vanish are generally expected to be non-analytic points for Schwinger functions.
Given a quantity $F(g)$ admitting a Borel resummable asymptotic expansion around $g=0$,
the region in the complex $g$ plane where the Borel reconstruction of the function is guaranteed to reproduce
the original function is given by a disk \cite{Sokal:1980ey} with a radius which is determined by the first singularities of $F(g)$ in the positive half-plane.
In our case the complex critical points are further away from the origin than the self-dual point.
This implies that the disk of minimal analyticity extends beyond the latter and allows us to explore (part of)
the strong branch when $\kappa<\kappa_*$.

\begin{figure}[t!]
  \includegraphics[width=0.487\textwidth]{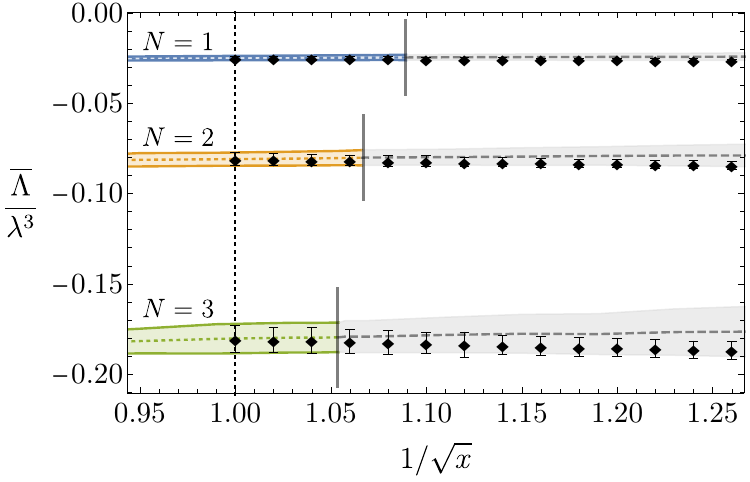}\quad
  \includegraphics[width=0.487\textwidth]{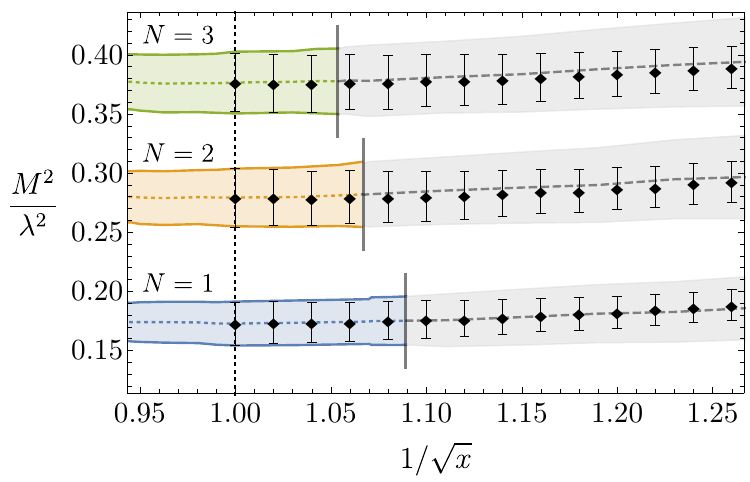}
  \centering
  \caption{The shifted vacuum energy $\overline{\Lambda} = \Lambda -\rho(\lambda)$ (with $\rho$ defined as in \eqref{eq:rhomu}) and the mass gap $M^2$ as a function of $\frac{1}{\sqrt{x}} = \frac{\sqrt{N+2}}{\pi} g$, for $N=1,\,2,\,3$.  The error bands and the central values (dashed lines) are obtained using conformal-Borel resummation.
    For any $N$ the self-dual point is at $x=1$. The points correspond to values obtained in the weak branch and mapped in the strong branch using the duality map  \eqref{eq:strongxd}.
    The vertical segment drawn on each band denotes the theoretical disk of analyticity: beyond that value the curves have been drawn in gray. To avoid overlapping of the curves we have applied an offset of $\Delta(M^2/\lambda^2) = (N-1)/10$ to the data in the right panel. In both panels $\kappa=5/2$.
  }
  \label{fig:Lambda_M2points}
\end{figure}

If self-duality is {\it assumed}, we can extract useful information on the asymptotic behavior of an observable $F(g)$ at strong coupling $g\rightarrow \infty$.
Let $F(g)$ be an observable with mass dimension $n$. After an appropriate rescaling we can write its Taylor expansion in the weak branch as
\begin{equation}
  F(g)  \sim m^n g^{k_0} f(g) \,, \quad \quad f(g) = 1+\sum_{k=1}^\infty c_k g^k\,, \quad g = \frac{\lambda}{m^{4-d}}\,,
\end{equation}
where $k_0\geq 0$ is the first non-vanishing order in perturbation theory. The $\sim$ is used because the series is only formal (asymptotic).
We consider both the $d=2$ and $d=3$ cases together, and for simplicity drop the tildes in $d=2$ on the couplings.
Self-duality implies
\begin{equation}
  F(g_w) = F(g_s) \quad \Rightarrow \quad
  g_w^{k_0 - \frac{n}{4-d}} f(g_w)= g_s^{k_0-\frac{n}{4-d}}  f(g_s)\,.
  \label{eq:Ogw=Ogs}
\end{equation}
In the limit $g_w \to 0$, we obtain the scaling at strong coupling from \eqref{eq:massS3d} as
\begin{equation}
  g_w^{-1} \sim \left( \log g_s \right)^{\frac{1}{d-1}}\,,
\end{equation}
which plugged into \eqref{eq:Ogw=Ogs} gives
\begin{equation}
  \lim_{g\rightarrow \infty} f(g) \sim g^{-k_0+\frac{n}{4-d}} (\log g)^\alpha\,, \quad \alpha = \frac{1}{d-1}\Big(\frac{n}{4-d} -k_0\Big)\,.
\end{equation}
Therefore the scaling of the observable $F(g)$ as $g\rightarrow \infty$  is
\begin{equation}
  F(g) \sim m^n g^s \left(\log g\right)^\alpha \,, \quad s = \frac{n}{4-d}\,.
  \label{eq:largeGo}
\end{equation}
Note that in general observables do not admit an analytic strongly coupled asymptotic Taylor expansion around infinity, due to the appearance of the logs.\footnote{Non-analytic
  expansions involving logarithms of the coupling have been invoked to cure IR divergences that appear with massless particles in 2d and 3d \cite{Jackiw:1980kv}.
  Interestingly enough, we see here how these log's automatically arise from the duality.}

We want to test the self-duality, so the scaling \eqref{eq:largeGo} will {\it not} be assumed. As a first indirect test of the duality, we find that the parameter $s$, which is fixed by the optimization procedure in our conformal-Borel resummations \cite{Serone:2018gjo}, is always close to the theoretical prediction \eqref{eq:largeGo} for both $\Lambda$ and $M^2$. Analogously, with the Pad\'e-Borel resummations we find that the best approximants $[p,q]$ satisfy the relation $p-q=s$.

We show in fig.~\ref{fig:Lambda_M2points} the vacuum energy $\Lambda$ and the mass gap $M^2$ as a function of $1/\sqrt{x}$ for different values of $N$
at $\kappa=5/2$. In order to take into account the vacuum energy shift, necessary to map it from the weak to the strong branch, we report the quantity $\overline{\Lambda} = \Lambda -\rho(\lambda)$, where $\rho$ is defined in \eqref{eq:rhomu}. In this way, $\overline{\Lambda}$ should have an extremum at the self-dual point
which,  in the variable $x$, is at $x=1$ for any value of $N$.
The black points in the figure correspond to values obtained in the weak branch and mapped in the strong branch using \eqref{eq:strongxd}.
The vertical segment drawn on each band denotes the disk of analyticity beyond which Borel resummation is not guaranteed to work.
Beyond that value, the curves have been drawn in gray. Fig.~\ref{fig:Lambda_M2points} gives us good evidence for the self-duality. Note in particular how $x=1$ is to a very good accuracy an extremum of both
$\overline{\Lambda}$ and $M^2$, as expected.
Interestingly enough, the agreement persists well beyond the disk of analyticity for both $\overline{\Lambda}$ and $M^2$.

\subsection{Scheme Dependence of Critical Couplings}
\label{sec:ReVEandM}

\begin{figure}[t!]
  \includegraphics[height=5.5cm]{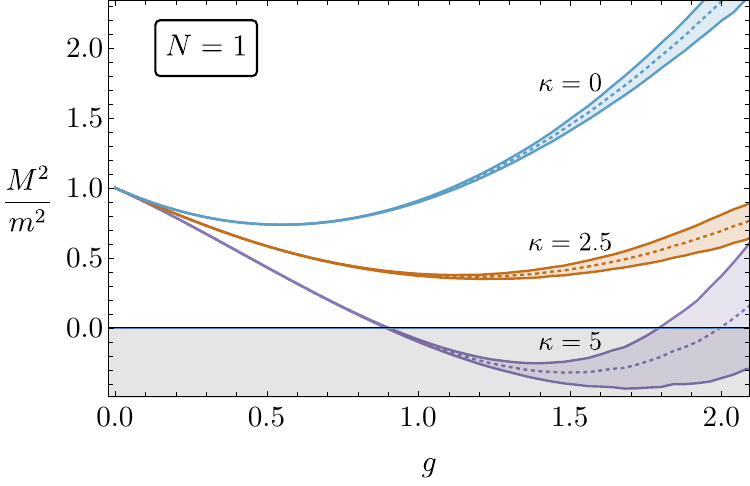}
  \qquad
  \centering
  \raisebox{0.33cm}{\includegraphics[height=5.2cm]{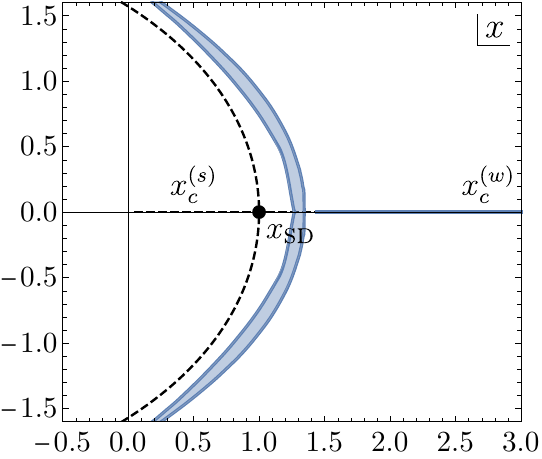}}
  \caption{(Left) The mass gap $M^2/m^2$ for $N=1$ at three different values of $\kappa$.
    (Right) The position of the critical coupling $x_c(\kappa)$ in the complex $x$ plane as $\kappa$ is varied for $N=1$. The blue bands are computed with conformal-Borel,
    the dashed black line is the analytic expectation from \eqref{eq:gcFunkappa}. }
  \label{fig:M2_kappa}
\end{figure}

In this subsection we determine how the critical coupling $g_c$ depends on the renormalization scheme.
We show in the left panel of fig.~\ref{fig:M2_kappa} $M^2$ as a function of $g$ for $N=1$ and different values of $\kappa$. As expected, the phase transition is not always visible and by increasing the value of $\kappa$ two zeros appear.  While the value of the first is in principle reliable and should be identified  with the weak critical coupling $g^{(w)}_c$, the same cannot be said for the second, since it is reached after the theory has passed a phase transition. Being $g=g_c^{(w)}$ a non-analytic point, Borel resummation is not guaranteed for $g>g_c^{(w)}$. For this reason we can only focus on the region where $g\leq g_c^{(w)}$.\footnote{It is however interesting to see that the analytic continuation of the Borel resummed mass gap $M^2$ for $g>g_c^{(w)}$ has a further zero, as expected from the self-duality of the theory
(see the purple band in the left panel of fig.~\ref{fig:M2_kappa}). The numerical accuracy of the resummation does not in any case allow us to determine the second zero accurately enough to possibly test if it is equal to $g_c^{(s)}$.}
The accuracy of the numerical resummations depends on $\kappa$ and only a limited range of optimal values of $\kappa$ (when the phase transition occurs) is expected. Indeed, as $\kappa$ decreases, the two critical couplings approach each other, and a general instability in the resummation procedure is expected and in fact does occur. On the other hand, if $\kappa$ increases, although the value of $g_c^{(w)}$ decreases, we are effectively in presence of large logs that spoil the validity of the perturbative expansion, as already noted in \cite{Sberveglieri:2019ccj}. We choose as optimal reference scheme $\kappa=5$ for any $N$.\footnote{The range of optimal values of $\kappa$ has a mild dependence on $N$, which can be neglected for low values of $N$.}

In the right panel of fig.~\ref{fig:M2_kappa} we plot the position of $g_c$ in the complex $g$-plane as $\kappa$ is varied and compare it with the analytic
prediction given by \eqref{eq:gcFunkappa}. The movement of $g_c$ as $\kappa$ varies is in fair agreement with the theoretical prediction, but it shows a small disagreement.
This discrepancy reflects a systematic slow convergence and low accuracy in the resummations for $\kappa>\kappa_*$.
In order to quantify it, we can compare the values of $g_c$ defined as the zero of $M^2$ and equivalently as the zero of the
function $L(g)=  (\partial_g \log M^2)^{-1}$. The function $L$ is useful because it can be used to extract the critical exponent $\nu$.
For example at $\kappa=5$, $N=1$, we find $g_c^{(M^2)}(\kappa=5) = 0.898(5)$ and  $g_c^{(L)}(\kappa=5) = 0.944(16)$. The two values are not in agreement and indicate the
presence of a systematic error which is not captured by our error estimate.
Similarly the accuracy in the determination of $\nu$ is  significantly lower than that found in the literature (see e.g. \cite{Guida:1998bx}) in the scheme of \cite{Parisi:1993sp}.
This lack of accuracy might be due to the presence of the self-duality and an analytic structure for observables more difficult to reconstruct numerically.

\begin{table}[t]
  \centering
  \begin{tabular}{c|c c c c}
    \hline
    Method             & $N=1$                       & $N=2$                         & $N= 4$                      \\\hline
    Lattice ${\rm MC}$ & 1.0670(17)\cite{Sun:2002cc} & 0.9509(5)\cite{Arnold:2001ir} & 0.8238(26)\cite{Sun:2002cc} \\\hline
    This work          & 1.08(3)                     & 0.94(2)                       & 0.80(2)                     \\\hline
  \end{tabular}
  \caption{Comparison of the (weak) critical coupling $g_c^\mathrm{MC}$  with the results of Lattice Monte Carlo computations, for 3d $O(N)$ models with $N=1,2,4$.}
  \label{tab:g_c}
\end{table}

The value of $g_c$ in $O(N)$ vector models has been computed in the past for $N=2$ and $N=1,4$ in \cite{Arnold:2001ir,Arnold:2001mu} and \cite{Sun:2002cc} respectively, using Lattice Monte Carlo methods. Very recently Hamiltonian truncation methods have been developed to study the $N=1$ theory \cite{EliasMiro:2020uvk,Anand:2020qnp}.
A comparison with our results is however still not available, because in \cite{EliasMiro:2020uvk} the extrapolation to infinite volume has not be taken and
in \cite{Anand:2020qnp} the use of light-cone quantization requires to work out the non-trivial map to pass to a covariant quantization. For this reason we restrict our comparison with the earlier results \cite{Arnold:2001ir,Arnold:2001mu,Sun:2002cc}.
These works report the value of $g_c$ in \MSb
at the scale $\mu=8\lambda$, which we denote by $g_c^{{\rm MC}}$. A direct  computation
at that scale is not possible, since our perturbative series will involve logarithms of $g$. However, we can access this value by using the exact one-loop running of $g_c(\kappa)$.
We get
\begin{equation}
  (g_c^{{\rm MC}})^{-2} = g_c^{-2}(\kappa)-\frac{N+2}{\pi^2} \log\Big(\frac{e^\kappa}{64 g_c^2(\kappa)} \Big)\,.
  \label{eq:gcLattice}
\end{equation}
The right hand side of \eqref{eq:gcLattice} should be independent of $\kappa$, but numerically a dependence on $\kappa$ remains.
We have computed $g_c^{(M^2)}$ and $g_c^{(L)}$ for a set of values of $\kappa \in [5, 6]$,  mapped them with \eqref{eq:gcLattice} and then taken an average value as our final estimate. In table \ref{tab:g_c} we compare these values of $g_c^\mathrm{MC}$ with those given by \cite{Arnold:2001ir,Arnold:2001mu,Sun:2002cc}. The values are in agreement, but with large errors on our side.

\section{Conclusions}
\label{sec:conclusions}

In this paper we have discussed the phase diagram of 3d $O(N)$ $\phi^4$ models using perturbation theory around the Gaussian fixed point.
In particular, we have reassessed the strong-weak Magruder duality in the classically unbroken phase and studied its renormalization scheme dependence.
Starting from the weak branch in perturbation theory, for certain schemes we encounter a critical coupling where the theory is gapless and a second-order phase transition
takes place. On the other hand, for other choices of schemes the theory is gapped and a pair of complex conjugate critical couplings appear.
In this case the weak and strong branches are no longer separated by a phase transition (for real values of masses and couplings)
and we can access the strong branch from the Gaussian fixed point. The phase transition is then no longer visible from the classically unbroken phase if one restricts to real parameters in the Lagrangian.
We have numerically verified these considerations by Borel resumming the perturbative series of the 3d $O(N)$ models for the vacuum energy and for the mass gap.

The merging of critical points is reminiscent of the fixed point annihilation advocated in \cite{Kaplan:2009kr} as a mechanism for loss of conformality
in QFT, see also \cite{Gorbenko:2018ncu,Gorbenko:2018dtm,Benini:2019dfy}. The fixed point annihilation described in those papers occur when parameters (such as the number of fields) in a family of critical theories are varied and so differ from the merging found in this paper, which is within a given theory when the renormalization scheme is varied.\footnote{Note however
  that a renormalization scheme dependence on the position of the critical points always occurs. It would be interesting to study more carefully the interplay between
  the position of fixed points determined by the parameters of the theory and by the renormalization scheme
  dependence of its couplings.} We see that merging of fixed points and complex CFTs (i.e. the two CFTs we would have at the two complex conjugate
complex values $x_c^{(w)}(\kappa)$ and $x_c^{(s)}(\kappa)$ for $\kappa < \kappa_*$) do not necessarily indicate an actual walking or first-order phase transition, but can be artefacts of the specific renormalization scheme chosen. In our case the complex CFTs should eventually correspond to the usual $O(N)$ symmetric CFTs, because they merely arise from a coupling constant redefinition.
It would be interesting however to better establish the correspondence, because it is not obvious if (and how) the CFT data of the two complex CFTs are in fact identical to those
of the ordinary unitary $O(N)$ symmetric CFTs.
The appearance and disappearance of fixed points makes also clear that a phase diagram of a theory is not universally determined, but
it depends on the renormalization scheme. For example, we see that according to fig.~\ref{fig:ChangRS} in the $N=1$ case the number of critical points
that occur in the entire range of the real squared mass parameter is either three or one, depending on the renormalization scheme.
The universal presence of a second-order phase transition could be argued from the fact  that this number modulo two is always one.\footnote{It would be nice to understand if this is associated to an index, in the spirit of \cite{Gukov:2016tnp}.}

The accuracy of our Borel resummations is worse than the one found in \cite{Serone:2018gjo} for the 2d $\phi^4$ theory, in contrast to
what happens in the scheme of \cite{Parisi:1993sp}, where results in 3d are more accurate than the ones in 2d.
We suspect that this might be due to the presence of the self-duality in the classically unbroken phase, which
gives rise to two fixed points and an analytic structure for observables more difficult to reconstruct numerically.

There are several open questions that deserve further study.
Two-dimensional  lattice spin systems with $\mathbb{Z}_2$ symmetry features Kramers-Wannier (KW) duality \cite{Kramers:1941kn} relating
the disordered and ordered phases. Being the $\phi^4$ theory the long-distance effective description of a $\mathbb{Z}_2$ lattice spin system,
it is reasonable to expect that KW duality persists when taking the continuum limit. The continuum version of KW duality is defined starting from the critical theory, i.e. the 2d Ising CFT point.
It is natural to conjecture that KW duality is closely related to Chang duality, and it would be interesting to find the precise map between the two. In particular, it would be nice to see how (if any) at finite volume a proper definition of Chang duality requires the presence of $\mathbb{Z}_2$ gauge fields, like KW does, see e.g. section 2 of \cite{Senthil:2018cru}.
Similarly, it would be interesting to see if there is a connection between Magruder duality of 3d $\phi^4$ theory and the continuum limit of the duality of a 3d Ising system with a $\mathbb{Z}_2$ lattice gauge theory (see e.g.~\cite{Savit:1979ny}).

Chang and Magruder dualities are crucially based on the super-renormalizability properties of $\phi^4$ theories in $d=2$ and $d=3$.
One can then try to derive similar dualities from more general super-renormalizable theories. The dualities so obtained would be in general only valid to all orders in perturbation theory,
but not non-perturbatively. In order to hope to have exact dualities, one should argue for the absence or decoupling of non-perturbative effects, like in the 2d $N=1$ and 3d $O(N)$ $\phi^4$ models in infinite volume studied in this paper. It is an interesting open question if there exists a generalization of this duality when gauge fields are added.
In particular, it would be interesting to see if a would-be Chang/Magruder-like duality of a gauged version of the 3d $O(2)$ model
can provide a ``UV completion" for particle-vortex duality \cite{Peskin:1977kp}.

\section*{Acknowledgments}

We thank Joan Elias-Mir\'o and Slava Rychkov for useful comments on the draft.
We also thank Nikhil Anand, Emanuel Katz, Zuhair U. Khandker and Matthew T. Walters for discussions and for having shared with us the draft of \cite{Anand:2020qnp} before publication.
GSp  would like to thank Gerard P. Lepage for the support given with the \texttt{vegas} python module.
GSb and MS are partially supported by INFN Iniziativa Specifica ST\&FI.

\appendix

\section{The Lambert $W$ Function}
\label{sec:Lambert}

\begin{figure}[t!]
  \includegraphics[width=0.55\textwidth]{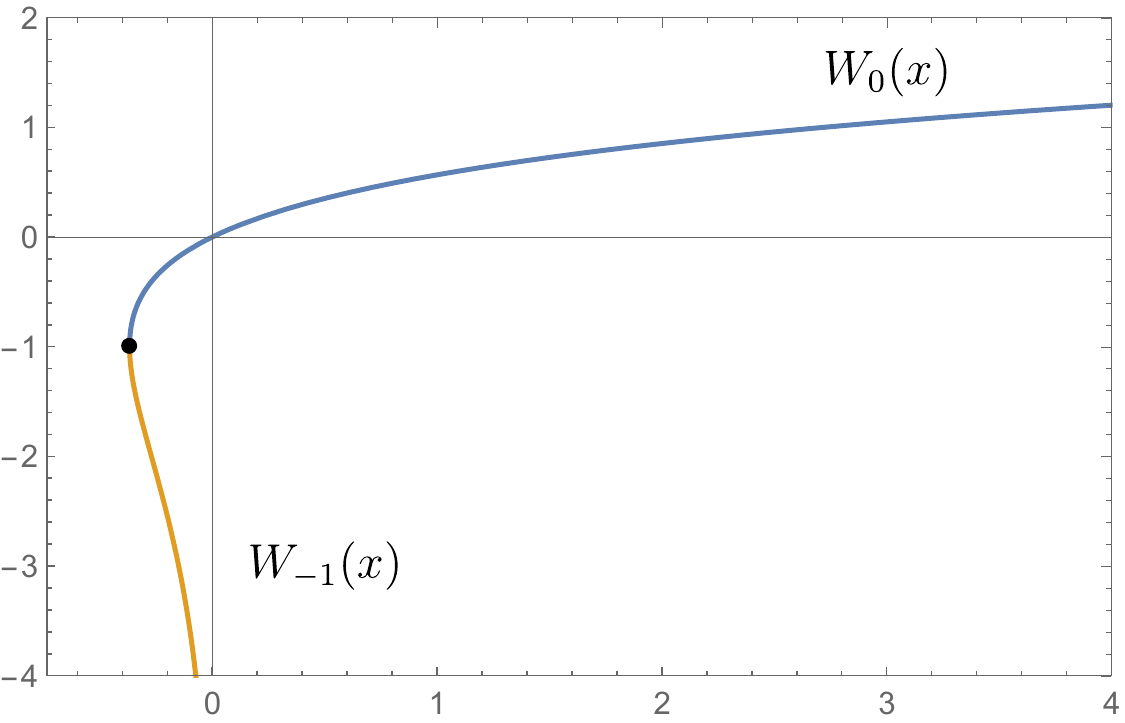}
  \centering
  \caption{The two branches of the Lambert function that take real values over $x$. The black dot corresponds to the branching point $x=-1/e$. }
  \label{fig:Lambert}
\end{figure}

Most of the results of this paper feature the Lambert function, so it is useful to review here some of its properties.
We refer the reader to \cite{COR96} for more details. The Lambert function is the function $W(x)$ that is obtained by inverting the relation
\begin{equation}
  w e^w = x \,.
\end{equation}
For large values of $w$ it behaves like the log function, but it deviates from it for small values. For $x>0$, $W(x)$ is monotonic, while
for $x<0$ it is double-valued, see fig.~\ref{fig:Lambert}. Over the reals, $W(x)$ has non-trivial support for $x\in[-1/e,\infty)$. As analytic complex function, $W(z)$ has an infinite number of branches, parametrized by an integer $k$. Only two branches, denoted by $W_0$ and $W_{-1}$, have real sections over $x$, see fig.~\ref{fig:Lambert}.
In all other branches $W_k(z)$, with $k\neq -1,0$, take complex values.  The function $W_0(z)$ is analytic at $z=0$ and it admits there the series expansion
\begin{equation}
  W_0(z) = \sum_{n=1}^\infty \frac{(-n)^{n-1}}{n!}z^n\,.
\end{equation}
The series above has a convergence radius equal to $1/e$. At $z=-1/e$ $W_0$ has a branch-cut singularity, where it branches into $W_1$ and $W_{-1}$.
Aside from $W_0$, all $W_k$ have a branch-cut at the origin and a logarithmic singularity at infinity. In particular, for any branch, we have
\begin{equation}
  \lim_{z\rightarrow \infty}  W_k(z) \approx \log z + 2i \pi k + {\cal O}(\log \log z)\,,
\end{equation}
and in particular for real $x$
\begin{equation}
  \lim_{x\rightarrow \infty}  W_0(x) \approx \log x - \log \log x + {\cal O}\Big(\frac{\log \log x}{\log x}\Big)\,.
  \label{eq:W0Largex}
\end{equation}
We will be mostly considering the branches $k=-1$ and $k=0$. A useful formula is
\begin{equation}
  \lim_{x\rightarrow 0^-} W_{-1}(x) =  \log (-x) +{\cal O}(\log(-\log( -x)))\,.
  \label{eq:Wm1x0}
\end{equation}
Another useful formula for $W$ is the following:
\begin{equation}
  \frac{d^n W(x) }{dx^n}= \frac{e^{-n W(x)} p_n(W)}{(1+W(x))^{2n-1}}\,, \quad n\geq 1\,,
  \label{eq:RRforL}
\end{equation}
where $p_n$ are polynomials of degree $n-1$ in $W$, defined by the recursion relation
\begin{equation}
  p_{n+1}(x) = - (n x+3n-1) p_n(x) +(1+x) p_n^\prime(x) \,, \quad \quad p_1(x) = 1\,.
\end{equation}
It is worth recalling a few QFT works where the Lambert $W$-function has appeared: in \cite{Gardi:1998qr} it has been shown that the two-loop
QCD beta-function can exactly be solved in terms of $W$ and studied the analyticity properties of the solution. In \cite{Bellon:2016mje} it has been shown that an infinite subset
of diagrams in the 4d SUSY massless  Wess-Zumino model can be resummed and leads to a beta-function and field anomalous dimension in terms of $W$.
A solution for the 2-point function for a non-commutative version of the 2d $\phi^4$ theory in terms of $W$ was found in \cite{Panzer:2018tvy}. More recently \cite{Borinsky:2020vae}
found that a subset of diagrams for the field anomalous dimensions in 4d massless Yukawa theory can be computed to all orders using a
truncation of the Schwinger-Dyson equations. The ansatz for the trans-series
associated with the known perturbative coefficients can be expressed in terms of W.

\section{Large $N$}

\label{sec:LL4D}

\begin{figure}[t!]
  \includegraphics[width=0.75\textwidth]{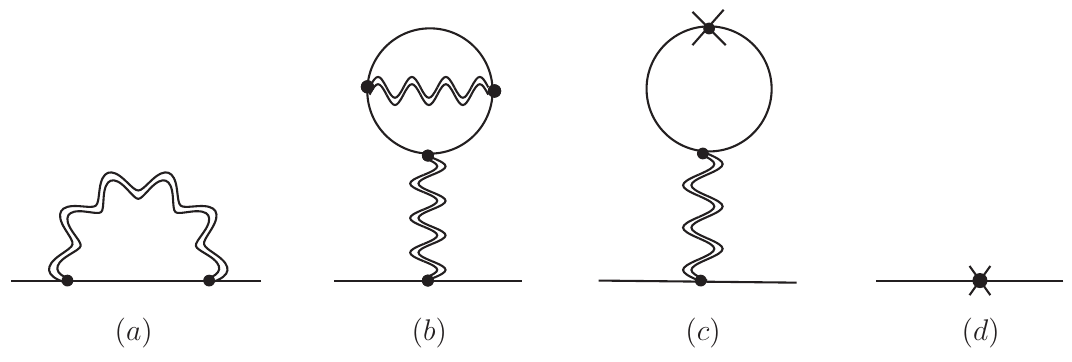}
  \centering
  \caption{Contributions of $o(N^{-1})$ to the two-point function $\langle \phi_i(-p) \phi_j(p)\rangle$. The counterterm depicted in (c) and (d) corresponds to $\hat\delta m^2$.}
  \label{fig:1oN}
\end{figure}

Large $N$ techniques are typically used in $O(N)$ models by taking $m^2=0$ and by going directly at the critical point, avoiding the problem of IR divergences.
In this way one can extract physical quantities such as scaling dimensions of the CFT operators, see e.g. section 2 of \cite{Fei:2014yja} for a clear and concise review.
In contrast, in this appendix we consider large $N$ of the massive $O(N)$ models, in line with the analysis in the main text.
In particular, we compute the vacuum energy $\Lambda = \Gamma^{(0)}$ and the mass gap $M^2 = \Gamma^{(2)}(p=0)$ at the first non-trivial leading order in large $N$ and to all orders in the coupling $\lambda$.
Although the diagrams surviving in the large $N$ limit are a small subset of the total and are not the hardest to determine, a comparison with large $N$ has been useful as a sanity check of the accuracy of the numerical evaluation of Feynman diagrams. We report here once again the euclidean action of the theory
\begin{equation}
  S = \int d^d x \Bigl[\frac12 (\partial_\mu \phi_i)^2+\frac 12 m_0^2 \phi_i^{2} + \lambda (\phi_i^{2})^2  + \rho_0 \Bigr]\,, \quad i=1,\ldots, N\,,
  \label{eq:S_appB}
\end{equation}
and we consider the large $N$-limit
\begin{equation}
  N \to \infty\,, \quad
  \lambda \to 0\,, \quad {\rm with } \quad
  \hat\lambda\equiv N \lambda = {\rm fixed}\,.
  \label{eq:largeNlimit}
\end{equation}
We define the renormalized parameters
\begin{equation}
  m_0^2 = m^2+ \delta m^2\,, \quad \quad
  \rho_0  = \rho+ \delta \rho\,,
\end{equation}
where
\begin{equation}
  \delta m^2  = \delta m^2_{(0)} + \frac 1N \delta m^2_{(1)} + o (N^{-2}) \,, \quad \quad \quad
  \delta \rho  = N \delta \rho_{(-1)} + \delta \rho_{(0)} + o (N^{-1})  \,,
  \label{eq:ctExp}
\end{equation}
and we choose a renormalization scheme where the vacuum energy counterterm $\delta\rho$  and the mass counterterm $\delta m^2$ exactly cancel the contributions in $\Lambda$ and $M^2$ up to order $\lambda^{d/(4-d)}$  and $\lambda^{2/(4-d)}$, respectively.\footnote{This is the generalization for any $d<4$ of what we denoted intermediate scheme in section \ref{subsec:coeff} of the main text. We have omitted in this appendix the subscript $I$ to avoid clutter.}
Introducing a Hubbard-Stratonovich auxiliary field $\sigma(x)$, we can rewrite $S$ as
\begin{equation}
  \hat S =  \int d^d x \Bigl[\frac12 (\partial_\mu \phi_i)^2+\frac 12 (m^2+\hat\delta m^2) \phi^{2}_i -\frac 12 \sigma^2+ \frac 12 \hat f \sigma  \phi^{2}_i+ \sigma \delta_T +\rho+\hat \delta \rho  \Bigr]\,.
  \label{eq:Ssigma}
\end{equation}
If we integrate out $\sigma$ we recover the action \eqref{eq:S_appB} provided we identify\footnote{The Gaussian integral in $\sigma$ is computed by analytic continuation from pure imaginary values, where the path integral converges.}
\begin{equation}
  \hat f = 2\sqrt{2 \lambda} \,, \quad \hat\delta \rho = \delta \rho - \frac{\delta_T^2}{2} \,, \quad \hat f \delta_T + \hat\delta m^2 = \delta m^2\,.
  \label{eq:parmatch}
\end{equation}
There is an arbitrariness in splitting the mass counterterm $\delta m^2$ in terms of $\delta_T$ and $\hat\delta m^2$. We choose
\begin{equation}
  \hat f \delta_T   = \delta m^2_{(0)} \,,  \quad \quad
  \hat  \delta m^2  = \frac{1}N \delta m^2_{(1)}\,,
  \label{eq:TadChoice}
\end{equation}
so that the tadpole counterterm $\delta_T$ for $\sigma$ completely cancels the radiatively induced tadpole at $o(N^0)$.
Let us first consider the 2-point function $\langle \phi_i(-p) \phi_j(p)\rangle = \Gamma_{ij}^{(2)}(p^2) \equiv \delta_{ij} \Gamma^{(2)}(p^2)$.
Since $\Gamma^{(2)}_{ij}$ is 1PI with respect to the $\phi_i$, but not with respect to $\sigma$, Feynman diagrams reducible when cutting a $\sigma$-propagator
should be considered. At $o(N^0)$ and $o(\hat \lambda)$ only one diagram contributes. In the chosen renormalization scheme
its contribution is canceled by $\delta_T$. The cancellation of tadpole-like graphs at $o(N^0)$ persists to all orders in $ \hat\lambda$,  so no contribution whatsoever arises at $o(N^0)$ in $\Gamma^{(2)}(p^2)$.
We now compute the $\langle \sigma \sigma \rangle$ propagator at $o (N^0)$. The relevant 1PI diagram is
\begin{equation}
  \raisebox{-.42\height}{\includegraphics[width=5cm]{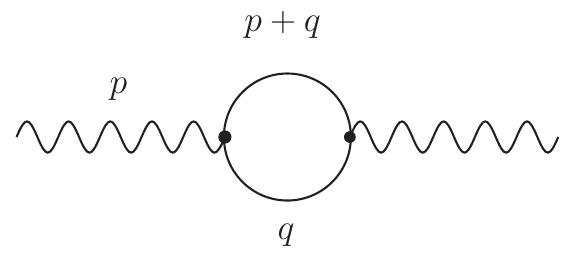}}
  = 4\hat\lambda \int\!\!\frac{d^dq}{(2\pi)^d}  \frac{1}{q^2+m^2}\frac{1}{(p+q)^2+m^2}  \,
  \equiv \, \hat\lambda \Pi_d(p^2) \,,
\end{equation}
where we used wavy lines for the field $\sigma$ along with the usual solid lines for the vector field $\phi_i$.
For $d<4$ the loop integral converges.
The resummation of the bubbles leads to the exact $o(N^0)$ propagator, which will be denoted by a double wavy line:
\begin{equation}
  \raisebox{-.55\height}{\includegraphics[width=3.8cm]{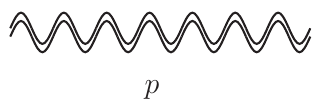}}  \equiv -\sum_{n=0}^\infty (-\hat\lambda\Pi_d(p^2))^n = - \frac{1}{1+\hat\lambda \Pi_d(p^2)}\,.
\end{equation}
We are now ready to study $\Gamma^{(2)}$ at $o(N^{-1})$. At this order 3 diagrams and the $\delta m^2_{(1)}$ counterterm contribute, see fig.~\ref{fig:1oN}.
Note that we also have $o(N^{-1})$ corrections to the $\sigma$ propagator, but these can enter in $\Gamma^{(2)}$ at this order only through tadpole graphs, and hence they vanish.
The divergences in graph $(a)$ arising from $n=0$ ($d<3$) or $n=1$ ($3\leq d < 4$) insertions of $\Pi_d$ in the expansion of the resummed propagator
are cancelled by the mass counterterm, so $(a)+(d)$  is finite.
Similar considerations apply for the graphs $(b)+(c)$.
We do not report the expressions for these graphs, that can be derived by standard manipulations.
Let us now consider the vacuum energy. In the chosen renormalization scheme, the $o (N)$ contributions to the vacuum energy are exactly canceled.\footnote{It is easy to see that the $o(N)$  counterterm $\delta \rho_{(-1)}$ in (\ref{eq:ctExp}) precisely cancels the term $\delta_T^2/2$ in (\ref{eq:parmatch}), which is also of order $o(N)$, so that the counterterm $\hat \delta \rho$ is $o(N^0)$.}
So the leading finite contribution arises at $o(N^0)$ and is given by a one-loop vacuum diagram of the exact  $o(N^0)$ $\sigma$-propagator.
Collecting the results above, we finally get
\begin{align}
  \Lambda           & =  \frac 12  \int\!\!\frac{d^dq}{(2\pi)^d}\log\Big(1+ \hat\lambda \Pi_d(q^2)\Big) + \delta\,, \label{app:rhoN}                                           \\
  M^2               & = m^2 +\frac{8\hat\lambda}N  \int\!\!\frac{d^dq}{(2\pi)^d} \sum_{n=2}^\infty \hat\lambda^{ n} \frac{(-\Pi_d(q^2))^n}{q^2+m^2} +(b)+(c)  + o (N^{-2}) \,, \\
  \Gamma^{(4)}(p=0) & =-\frac{24\hat\lambda}{N} \frac{1}{1+\hat\lambda \Pi_d(0)} +o (N^{-2})\,.
\end{align}
We have also reported the leading order  expression of the 4-point 1PI function $\Gamma^{(4)}$, which is $o(N^{-1})$, and is trivially given by tree level diagrams only.
There is no need to keep track of the form of the counterterm $\delta$ appearing in \eqref{app:rhoN}, because in our scheme it is equal and opposite to the first divergent terms arising from the loop integral when expanded in powers of $\hat\lambda$. The form of $\Pi_d(q^2)$ and more explicit expressions for
$M^2$ will be given below for the specific $d=2$ and $d=3$ cases.
In what follows it will be useful to use dimensionless quantities and rewrite
\begin{equation}
  \hat\lambda \Pi_d(q^2) \equiv \hat g \,U_d(y), \quad \quad \hat g \equiv \frac{\hat\lambda}{m^{4-d}}\,, \quad \quad y\equiv \frac{q^2}{4m^2}\,.
\end{equation}

\subsection{$d=2$}
We specialize here to the $d=2$ case.
Working out the contributions from graphs $(b)+(c)$ in fig.~\ref{fig:1oN}, we obtain the following expression for $M^2$:
\begin{equation}
  \frac{M^2 }{m^2}  = 1-\frac{8\hat g}{\pi N} \int_0^\infty \! dy \sum_{n=1}^\infty (-\hat g \,U_2(y))^n \bigg(\frac{1}{1+4y} -
  \frac{\hat g}{1+ \frac{\hat g}{\pi}}  V_2(y) \bigg)  + o (N^{-2}) \,,
\end{equation}
where
\begin{equation}
  U_2(y) = \frac{1}{\pi} \frac{\log(\sqrt{y}+\sqrt{1+y})}{\sqrt{y(1+y)}}\,,
  \qquad
  V_2(y) \equiv \frac{1}{4\pi} \frac{\sqrt{y(y+1)} +  \arctanh \Big(\sqrt{\frac{y}{1+y}}\Big)}{\sqrt{y} (1+y)^{3/2}}\,.
\end{equation}
We report below the numerical values for the first coefficients in an expansion in $\hat g$ of $\Lambda$ and $M^2$:
\begin{align}
  \Lambda  =                     & \; - 0.016961 \hat g^2 + 0.0015425 \hat g^3 - 0.00023173 \hat g^4 +o (\hat g^5) + 	o (N^{-1})		\,,		\nn \\
  \frac{M^2 }{m^2}  =            & \;  1 +\frac 1N \Big( \hat g(
  -0.47497 \hat g^2 + 0.23046 \hat g^3 - 0.090670 \hat g^4 +o (\hat g^5) \Big)+ 	o (N^{-2})	\,,		\nn                                      \\
  \frac{\Gamma^{(4)}(0) }{m^2} = & -\frac{24 \hat g}{N }\frac{1}{1+\frac{\hat g}{\pi}}	 + 	o (N^{-2})	\label{eq:2dresults}	\,.
\end{align}
It is known that generally the large order behavior of the large $N$ coupling expansion, at given order in $N$, is convergent.
The above results are in agreement with this expectation. From a numerical exploration we find that the series in $\hat g$ for $\Lambda$ at $o (N^0)$ and  $\Gamma^{(2)}$  at $o(N^{-1})$ \ are convergent, with a radius of convergence equal to $\pi$. This is in agreement with the radius of convergence of $\Gamma^{(4)}$ that is manifest from its analytic form at $o(N^{-1})$.

\subsection{$d=3$}

Proceeding as above for the $d=3$ case, we get the following
expression for $M^2$:
\begin{align}
  \frac{M^2 }{m^2}  = & \;  1 - \frac{\hat g^3 \log \big(\frac 43\big)}{N \pi^3(1+ \frac{\hat g}{2\pi})}+\!\frac{2\hat g}{N\pi^2}  \int_0^\infty\!\!\!\! \!dy \sqrt{y} \sum_{n=2}^\infty (-\hat g U_3(y))^n
  \bigg( \frac{8}{4y+1} - \frac{\hat g}{\pi(1+ \frac{\hat g}{2\pi})} \frac{1}{y+1} \bigg)
  + o (N^{-2})\,, \nn
\end{align}
where the function $U_3(y)$ is given by
\begin{equation}
  U_3(y) = \frac{1}{4\pi} \frac{\arccot \big(\frac{1}{2y}\big)}{y}\,.
\end{equation}
The numerical values for the first coefficients in an expansion in $\hat g$ read
\begin{align}
  \Lambda  =                   & \;-0.000073108 \hat g^4 + 3.4816\times 10^{-6} \hat g^5  +o (\hat g^6) + 	o (N^{-1})		\,,		\nn           \\
  \frac{M^2 }{m^2}  =          & \;  1  +\frac 1N \Big( 0.023840 \hat g^3 - 0.0053959 \hat g^4 +o (\hat g^5) \Big)+ 	o (N^{-2})	\,,		\nn \\
  \frac{\Gamma^{(4)}(0) }{m} = & -\frac{24\hat g}{N }\frac{1}{1+\frac{\hat g}{2\pi}}	 + 	o (N^{-2})		\label{eq:3dresults}	\,.
\end{align}
Like in the $d=2$ case, the series in $\hat g$ for $\Lambda$ at $o (N^0)$ and  $\Gamma^{(2)}$  at $o(N^{-1})$ are convergent, with a radius of convergence equal to $2\pi$. This is in agreement with the radius of convergence of $\Gamma^{(4)}$ that is manifest from its analytic form at $o(N^{-1})$.

\section{Vacuum Energy Renormalization in $d=3$}
\label{app:MVEd3}

In the following we derive the counterterm for the vacuum energy in the \MSb scheme needed to establish the duality of the theory.

First, we recall the determination of the mass counterterm $\delta m^2$.
\begin{figure}[t]
  \centering
  \footnotesize
  \begin{tabular}{>{\centering}p{2.5cm}p{0.5cm}>{\centering}p{2.5cm}p{0.5cm}p{2.5cm}<{\centering}}
    $\Sigma_1$      &        & $\Sigma_{2a}(k)$     &  & $\Sigma_{2b}$          \\
    \includegraphics[height=\diagh]{Figs/2pt_o1.png}
                    & \qquad &
    \includegraphics[height=\diagh]{Figs/2pt_o2_a.png}
                    & \qquad &
    \includegraphics[height=\diagh]{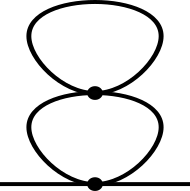}
    \\
    $4\lambda(N+2)$ &        & $-32\lambda^2 (N+2)$ &  & $-16\lambda^2 (N+2)^2$
  \end{tabular}
  \caption{The two-point diagrams up to 2 loops together with the multiplicity factors.}
  \label{fig:2pt_diags}
\end{figure}
Within dimensional regularization only the \emph{sunset} diagram has a pole in $\epsilon=d-3$ and contributes to the mass counterterm $\delta m^2$.
Below we give the explicit expressions for the three diagrams in fig.~\ref{fig:2pt_diags}:
\begin{align*}
  \Sigma_{1}
                 & = - \lambda m \frac{N+2}{\pi} \,,
  \\
  \Sigma_{2a}(k) & =  -\lambda^2 \frac{N+2}{\pi^2}   \left[\frac{1}{\epsilon} + 3 + \log \frac{\mu^2}{9 m^2} -\log \left(1+\frac{k^2}{9m^2}\right)-\frac{6 m}{|k|} \arctan\left(\frac{|k|}{3m}\right) \right]\,,
  \\
  \Sigma_{2b}    & = \lambda^2 \frac{(N+2)^2}{2\pi^2} \,.
\end{align*}
Hence we find
\begin{equation}
  \delta m^2 = \frac{\lambda^2}{\epsilon} \frac{N+2}{\pi^2} \,.
\end{equation}

\begin{figure}[t]
  \centering
  \footnotesize
  \begin{tabular}{
    >{\centering}p{2.9cm}
    >{\centering}p{2.9cm}
    >{\centering}p{2.9cm}
    >{\centering}p{2.9cm}
    p{2.9cm}<{\centering}
    }
    $\Upsilon_{0}$           & $\Upsilon_{1}$                                     & $\Upsilon_{2a}$         & $\Upsilon_{2b}$      & $\Upsilon_{2c}$                      \\
    \includegraphics[height=\diagh]{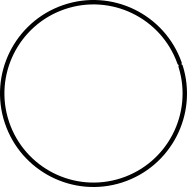}
                             & \includegraphics[height=\diagh]{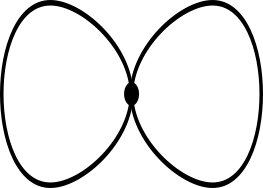}
                             & \includegraphics[height=\diagh]{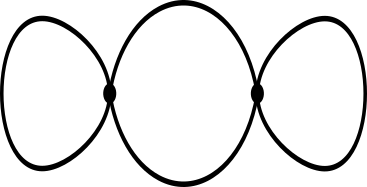}
                             & \includegraphics[height=\diagh]{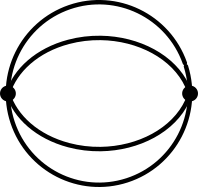}
                             & \includegraphics[height=\diagh]{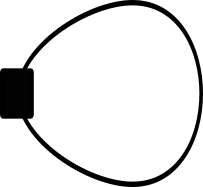}
    \\
    $N/2$                    & $\lambda N(N+2)$                                   & $-4\lambda^2 N(N+2)^2$  & $-4\lambda^2N(N+2)$  & $N/2$                                \\
    \\[0.1cm]
    $\Upsilon_{3a}$          & $\Upsilon_{3b}$                                    & $\Upsilon_{3c}$         & $\Upsilon_{3d}$      & $\Upsilon_{3e}$                      \\
    \includegraphics[height=\diagh]{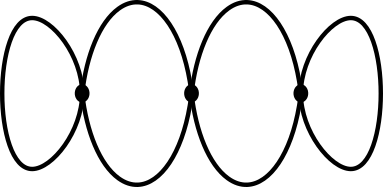}
                             & \includegraphics[height=\diagh]{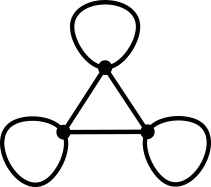}
                             & \includegraphics[height=\diagh]{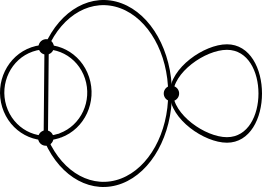}
                             & \includegraphics[height=\diagh]{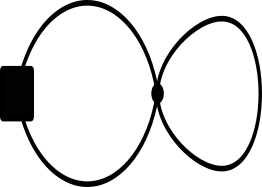}
                             & \includegraphics[height=\diagh]{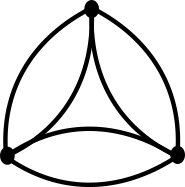}
    \\
    $16 \lambda^3 N (N+2)^3$ & $\frac{32}{3} \lambda^3 N (N+2)^3$                 & $64\lambda^3 N (N+2)^2$ & $- 2\lambda N (N+2)$ & $\frac{32}{3} \lambda^3 N(N+2)(N+8)$ \\
  \end{tabular}
  \caption{The zero-point diagrams up to 4 loops together with the multiplicity factors. The filled black squares represent factors of $\delta m^2$.}
  \label{fig:0pt_diags}
\end{figure}

Secondly, we turn to the determination of the vacuum energy counterterm $\delta \rho$. Since the divergences in the vacuum energy can be found up to four loops we have explicitly computed the diagrams in fig.~\ref{fig:0pt_diags} within dimensional regularization and we report their values below.
The contributions at order zero and one are finite within dimensional regularization and give
\begin{equation*}
  \Upsilon_{0} = - m^3 \frac{N}{12\pi}\,,
  \qquad
  \Upsilon_{1} = \lambda m^2 \frac{N(N+2)}{16\pi^2}  \,.
\end{equation*}
At order two we find two diagrams giving $1/\epsilon$ poles that cancel out:
\begin{align*}
  \Upsilon_{2a} & = - \lambda^2 m \frac{N(N+2)^2}{32\pi^3}\,,
  \\
  \Upsilon_{2b} & = \lambda^2 m \frac{N(N+2)}{8\pi^3}\left[\frac{1}{\epsilon } -\frac32 \log \frac{ m^2}{\mu^2} + 4  - 5\log 2 \right]\,,
  \\
  \Upsilon_{2c} & =
  - \lambda^2 m \frac{N(N+2)}{8\pi^3} \left[ \frac{1}{\epsilon} - \frac12 \log\frac{m^2}{\mu^2} +1 - \log 2  \right] \,.
\end{align*}
At order three the poles given by the diagrams $\Upsilon_{3c}$ and $\Upsilon_{3d}$ cancel out, leaving one divergent contribution from $\Upsilon_{3e}$ only:
\begin{align*}
  \Upsilon_{3a} & =
  \lambda^3 \frac{N(N+2)^3}{64\pi^4} \,.
  \\
  \Upsilon_{3b} & =
  -\lambda^3 \frac{N(N+2)^3}{192\pi^4} \,,
  \\
  \Upsilon_{3c} & =
  -\lambda^3 \frac{N(N+2)^2}{16\pi^4}\left( \frac{1}{\epsilon} -2\log\frac{m^2}{\mu^2} + 2 -6\log 2 \right)
  \,,
  \\
  \Upsilon_{3d} & =
  \lambda^3 \frac{N(N+2)^2}{16\pi^4} \left[\frac{1}{\epsilon}- \log\frac{m^2}{\mu^2}  +  1  - 2 \log 2 \right]\,,
  \\
  \Upsilon_{3e} & =
  \lambda^3 \frac{N(N+2)(N+8)}{384\pi^2}\left( \frac{1}{\epsilon}-2\log\frac{m^2}{\mu^2} + 1-2\log 2-\frac{42 \zeta(3)}{\pi^2} \right)
  \,.
\end{align*}
Therefore the counterterm $\delta \rho$ is determined as
\begin{equation}
  \delta \rho = -\frac{\lambda^3}{\epsilon} \frac{N(N+2)(N+8)}{384\pi^2} \,,
\end{equation}
which implies
\begin{equation}
  \beta_\rho =
  -4\lambda^3 \frac{N(N+2)(N+8)}{384\pi^2} \,,
  \qquad
  \rho(\mu) = \rho(m)+\lambda^3 \frac{N(N+2)(N+8)}{384\pi^2}\,2 \log \frac{m^2}{\mu^2}\,.
  \label{eq:rhomu}
\end{equation}
The self-duality is then obtained by mapping the parameters between the theory at the scale $m$ and the theory at the scale $\mt$. In other words we find the scale $\mu = \mt$ such that $m^2(\mt) = \mt^2$. From \eqref{eq:rhomu} we find the constant contribution to the vacuum energy $\rho(\mt)$ that one has to take into account in order to completely match the two theories.

\section{Series Coefficients for $\Lambda$ and $M^2$}
\label{app:Coeff3d}

In this appendix we report the coefficients for the series expansion of the vacuum energy $\Lambda$ and of the mass gap $M^2$ in the \MSb scheme (i.e. at $\mu=m$ or equivalently $\kappa=0$)
obtained as explained in section \ref{subsec:coeff}. The numerical coefficients appearing without error have been computed to a higher accuracy and have been truncated here to nine relevant digits.

\begin{scriptsize}
  \begin{equation}
    \begin{split}
      \frac{\Lambda-\rho}{m^3}
      =
      &  -\frac{N}{12\pi}
      + g \frac{N(N+2)}{16\pi^2}
      - g^2  \frac{N(N+2)}{8\pi^3}
      \left(
      \frac{N+2}{4} -3 + 4\log 2
      \right)\\
      & - g^3 \frac{N(N+2)}{384\pi^4}
      \left[
        (N+8) \left(42 \zeta (3) - \pi^2 + 2\pi^2 \log2 \right) - 24 (N+2) (4\log2 -1) - 4 (N+2)^2
        \right]
      \\
      &- g^4 \bigg[
        0.01030303(80\pm14)N
        +0.00737580(23\pm10) N^2
        +0.001384239(45\pm16) N^3\\ &\hspace*{1cm}
        +0.000148813598N^4
        +\frac{N^5}{512\pi^5}
        \bigg]
      \\
      & + g^5 \bigg[
        0.0017438(5\pm8)N
        -7.9(7\pm9)\times10^{-6}N^2
        - 0.0005433(15\pm31)N^3\\ &\hspace*{1cm}
        - 0.00004632(08\pm26)N^4
        + 2.68116618 \times10^{-6} N^5
        \bigg]
      \\
      & - g^6 \bigg[
        0.0034810(7\pm8) N
        +0.0020412(1\pm9) N^2
        +0.0000299(7\pm4) N^3
        - 0.0000603(41\pm10) N^4\\ &\hspace*{1cm}
        - 6.34(6\pm8) \times 10^{-7} N^5
        - \left(3.31589818 \times10^{-7}\right) N^6
        - \frac{N^7}{12288\pi^7}  \label{eq:Gamma0_3d}
        \bigg]
      \\
      & + g^7 \bigg[
        0.0046384(0\pm9) N
        + 0.003119(02\pm11) N^2
        + 0.0002951(2\pm6) N^3
        - 0.0000433(97\pm17) N^4 \\ &\hspace*{1cm}
        + 5.82(75\pm22) \times 10^{-6} N^5
        + 5.99(86\pm13) \times 10^{-7} N^6
        - 3.20181996\times10^{-8} N^7
        \bigg]
      \\
      & - g^8 \bigg[
        0.00705(0\pm7) N
        + 0.00497(6\pm9) N^2
        + 0.00050(6\pm4) N^3
        - 0.00012(47\pm10) N^4\\ &\hspace*{1cm}
        - 4.(67\pm11) \times10^{-6}N^5
        + 1.31(4\pm7) \times10^{-6}N^6
        - 2.5(19\pm11) \times10^{-8}N^7 \\ &\hspace*{1cm}
        + 1.846631(00\pm33) \times10^{-9} N^8
        + \frac{N^9}{131072\pi^9}
        \bigg]
      \,.
    \end{split}
  \end{equation}
  \begin{equation}
    \begin{split}
      \frac{M^2}{m^2}
      =\: & 1 - g \frac{N+2}{\pi } + g^2 \frac{(N+2) (N+4\log 3)}{2 \pi ^2} 
      + g^3 \bigg[
        0.254293918
        + 0.0394597748 N
        - 0.0519064757 N^2
        - \frac{N^3}{8\pi^3}
        \bigg]
      \\
      & - g^4 \bigg[
        0.3078241 (2\! \pm \! 5 )
        +0.1706010(23 \pm 33) N
        +0.00128178(4 \pm 5) N^2
        -0.00353134874 N^3
        \bigg]
      \\
      & + g^5 \bigg[
        0.383625   (87 \pm  23 )
        + 0.249106 (00 \pm  22 ) N
        + 0.0219413 (0 \pm  8 ) N^2 \\ &\hspace*{1cm}
        - 0.00299454 (5 \pm  9 ) N^3
        + 0.000230093158N^4
        +\frac{N^5}{128 \pi ^5}
        \bigg]
      \\
      &-g^6 \bigg[
        0.557150 (5 \pm  8)
        +0.38254 (40 \pm  10 ) N
        + 0.038154   (1 \pm  5 ) N^2
        - 0.007604   (05 \pm  12 ) N^3 \\ &\hspace*{1cm}
        -0.0002343 (00 \pm  11 ) N^4
        +0.0000550745909 N^5  \label{eq:Gamma2_3d}
        \bigg]
      \\
      &+ g^7 \bigg[
        0.97639(2\pm5)
        + 0.73243(8\pm7) N
        + 0.10656(30\pm35) N^2
        -0.001202(74\pm10) N^4  \\ &\hspace*{1cm}-0.010582(1\pm8)N^3
        +  0.00009681(5\pm7)N^5
        - 1.93299923(26\pm35) \times 10^{-6} N^6
        -\frac{N^7}{1024\pi^7}
        \bigg]
      \\
      &-g^8 \bigg[
        1.9235(06\pm21)
        +1.5462(72\pm32) N
        +0.2816(17\pm19) N^2
        - 0.01295(5\pm6) N^3 -0.003382(5\pm9) N^4 \\ &\hspace*{1cm}
        +0.0002399(6\pm7) N^5
        +9.18(54\pm28) \times 10^{-6} N^6  
        -1.08546706(47\pm15) \times 10^{-6} N^7
        \bigg]\,.
    \end{split}
  \end{equation}
\end{scriptsize}
Recall that we have normalized the vacuum energy by taking $\rho=\rho(m)=0$.

\bibliographystyle{JHEP}
\bibliography{Refs}

\end{document}